\documentclass[journal]{IEEEtran}

\usepackage[T1]{fontenc} 
\usepackage{cite}
\usepackage{subfigure}
\usepackage{graphicx}
\usepackage{amsmath}
\usepackage{mathtools}
\usepackage{textcomp}
\usepackage{etoolbox}
\usepackage{comment}
\usepackage{enumitem}
\usepackage{booktabs}
\usepackage{multirow} 
\usepackage{epstopdf}
\usepackage{gensymb}
\usepackage{array}
\usepackage{wrapfig}
\usepackage[flushleft]{threeparttable}
\usepackage{xcolor}
\usepackage{soul}

\newcommand{\innermid}{\nonscript\;\delimsize\vert\nonscript\;}
\newcommand{\activatebar}{
  \begingroup\lccode`\~=`\|
  \lowercase{\endgroup\let~}\innermid 
  \mathcode`|=\string"8000
}
\usepackage{amsthm,amssymb}
\usepackage{setspace}
\usepackage{rotating,booktabs}
\newcommand{\subparagraph}{}
\usepackage[compact]{titlesec}
\usepackage{hyperref}
\hypersetup{
    colorlinks=false,
}

\usepackage{bm}
\usepackage{tikz}
\usetikzlibrary{shapes,arrows}
\def\tablename{TABLE}
\begin{document}

\title{Study of 3D Virtual Reality Picture Quality}

\author{Meixu Chen,~\IEEEmembership{Student Member,~IEEE,}
        Yize Jin, Todd Goodall, Xiangxu Yu,
        and Alan C. Bovik,~\IEEEmembership{Fellow,~IEEE}
\thanks{M. Chen, Y. Jin, X. Yu, and A. C. Bovik are with the Department of Electrical and Computer Engineering, University of Texas at Austin, Austin, USA (e-mail: chenmx@utexas.edu; yizejin@utexas.edu; yuxiangxu@utexas.edu; bovik@ece.utexas.edu). T. Goodall is with Facebook Reality Labs, Facebook, California, USA (email: Todd.Goodall@oculus.com).}}


\maketitle

\begin{abstract}
Virtual Reality (VR) and its applications have attracted significant and increasing attention. However, the requirements of much larger file sizes, different storage formats, and immersive viewing conditions pose significant challenges to the goals of acquiring, transmitting, compressing and displaying high quality VR content. Towards meeting these challenges, it is important to be able to understand the distortions that arise and that can affect the perceived quality of displayed VR content. It is also important to develop ways to automatically predict VR picture quality. Meeting these challenges requires basic tools in the form of large,  representative subjective VR quality databases on which VR quality models can be developed and which can be used to  benchmark VR quality prediction algorithms. Towards making progress in this direction, here we present the results of an immersive 3D subjective image quality assessment study. In the study, 450 distorted images obtained from 15 pristine 3D VR images modified by 6 types of distortion of varying severities were evaluated by 42 subjects in a controlled VR setting. Both the subject ratings as well as eye tracking data were recorded and made available as part of the new database, in hopes that the relationships between gaze direction and perceived quality might be better understood. We also evaluated several publicly available IQA models on the new  database, and also report a statistical evaluation of the performances of the compared IQA models. 
\end{abstract}

\IEEEpeerreviewmaketitle
\section{Introduction}
\IEEEPARstart{V}{irtual} Reality (VR) and its applications have evolved quickly in recent years since the launches of popular head-mounted consumer displays like the Oculus Rift, HTC Vive and PlayStation VR.  Revenues from VR apps, gaming and video reached nearly 4 billion dollars in 2017 and are
expected to soar more than fivefold
by 2022\cite{pwcreport}. Given the recent availability of cheaper standalone headsets, like the Oculus Go, and the development of faster and more reliable 5G wireless networks, the installed base of headsets is expected to grow substantially. VR is being used in an increasing variety of consumer applications, including gaming, 360-degree image and video viewing, and visually immersive education. Websites like Youtube, Facebook and Netflix now support 360 image and video viewing and are offering a variety of online resources, further stimulating more consumer participation in VR.

Unlike traditional images, VR images are usually captured using a 360 camera equipped with multiple lenses that capture the entire 360 degrees of a scene. For example, the Samsung Gear 360 VR Camera is a portable consumer VR device with 180$\degree$ dual lenses that can capture images of resolution up to 5472 $\times$ 2736. The recent Insta360 Titan is a professional 360 camera with eight 200$\degree$ fisheye lenses that can capture both 2D and 3D images of resolution up to 11K. After the images are captured simultaneously by separate lenses, they are stitched together to generate a spherical image. The spherical image is usually stored in equirectangular projection format. Stereoscopic images are usually stored in an over-under equirectangular format,  where the left image is on top and the right one is on the bottom. Unlike traditional viewing conditions where people watch images and videos on flat-panel computer and mobile displays, VR offers a more immersive viewing environment. Since the image can cover the entire viewing space, users are free to view the image in every direction. Usually, only a small portion of the image is displayed as they gaze in any given direction, so the content that a  user sees is highly dependent on the spatial distribution of image content, the object being fixated on, and the spatial distribution of visual attention. The free-viewing of high resolution, immersive VR implies significant data volume, which leads to challenges when storing, transmitting and rendering the images which can affect the viewing quality. Therefore, it is important to be able to analyze and predict the perceptual quality of immersive VR image. 

Both subjective and objective tools are needed to understand and assess immersive VR images quality. Subjective VR image quality assessment (VR-IQA) is a process whereby the quality of VR images is rated by human subjects. The collected opinion scores supply the gold standard ground truth on which predictive models can be designed or tested. To our knowledge, there are only a few existing VR databases that include subjective measurements. Most only include traditional distortions such as image compression artifacts, Gaussian noise and Gaussian blur, but fail to capture distortions that are unique to panoramic  VR (2D and 3D) images. Towards advancing progress in this direction, we have created a more comprehensive database that both includes traditional image distortions as well as VR-specific  stitching distortions. We also include  eye tracking data that was obtained during the subjective study. The new LIVE 3D VR IQA Database will be made publicly available for free to facilitate the development of 2D and 3D VR IQA models by other research groups. The database can be accessed at \url{ http://live.ece.utexas.edu/research/VR3D/index.html}. 

The rest of the paper is organized as follows. 
Section \ref{section2} briefly introduces current progress on objective and subjective VR-IQA research.
Section \ref{section3} describes the details of the subjective study.
Section \ref{section4}  discusses data analysis of the subjective study results.
Section \ref{section5} analyzes the performances of a variety of objective IQA models on the new database, and Section \ref{section6} concludes the paper with some discussion of future research directions.

\section{Related works}
\label{section2}
\subsection{Subjective Quality Assessment}
Although it dates from at least as early as the 1970's, VR has been a topic of considerably renewed interest since the appearance of the Oculus Rift DK1. Viewing images in VR gives a more realistic and immersive viewing experience arising from the large field of view, 360\textdegree{} free navigation and the sense of being within a virtual environment. 

However, the immersive environment incurs a significant computational cost. VR images and videos are much larger than traditional planar images displayed on computer or TV screens and require much higher transmission bandwidth and significantly greater computational power. These demands are hard to meet, often at the cost of errors in capture, transmission, coding, processing, synthesis, and display. These errors often degrade the visual quality by introducing blur, blocking, transmission, or stitching artifacts. Therefore, developing algorithms for the automated quality assessment of VR images will help enable the future development of VR technology. 
Developing these algorithms requires subjective data for design, testing, and benchmarking. There are many widely used image quality databases, such as the LIVE Image Quality Assessment Database\cite{sheikh2006statistical}, the TID2013 database\cite{ponomarenko2015image},  CSIQ\cite{larson2010categorical}, and the LIVE In-the-Wild Challenge Database\cite{ghadiyaram2016massive}. These databases embody a wide variety of image distortions, but they were built for the purpose of studying traditional "framed" 2D images and are not suitable for building or testing algorithms designed to assess VR images.

Recently, there has been increasing interest in developing VR databases, and progress has been made in this direction. Duan \textit{et al.}\cite{duan2017ivqad} developed an immersive video database, containing downsampling and MPEG-4 compression distortions. In \cite{upenik2016testbed}, Upenik \textit{et al.} introduced a mobile testbed for evaluating immersive images and videos and an immersive image database with JPEG compression. Sun \textit{et al.}\cite{sun2017cviqd} constructed the Compression VR Image Quality Database (CVIQD), which consists of 5 reference images and corresponding compressed images created using three coding technologies: JPEG, H.264/AVC and H.265/HEVC. An omnidirectional IQA (OIQA) database established by Duan \textit{et al.}\cite{duan2018perceptual} includes four distortion types, JPEG compression, JPEG2000 compression, Gaussian blur and Gaussian noise. This database also includes head and eye tracking data that compliment the objective ratings. Another database that both includes head and eye movement data, VQA-OV, was proposed in \cite{Li:2018:BGV:3240508.3240581}. This database includes impairments from both compression and map projection. Xu \textit{et al.}\cite{xu2017subjective} also established a database with viewing direction data on immersive videos. However, most of the available VR databases only include distortions that occur in planar images, but without distortions that are specific to VR, such as stitching. Moreover, newer compression methods such as VP9 are relevant to the encoding of VR images, and these other compression methods are likely to play a substantive role in the future. Furthermore, amongst all the existing VR databases, to the best of our knowledge there are no 3D VR image quality databases as of yet.

\subsection{Objective Quality Assessment}
Both MSE and PSNR were long used as the basic way to assess image and video quality prior to the appearance of modern image objective quality assessment (IQA) methods. These IQA methods can be classified as: full reference (FR-IQA), reduced reference (RR-IQA) , or no reference (NR-IQA). Full reference IQA is appropriate when an undistorted, pristine reference image is available. Reduced reference IQA models only require partial reference information, and no reference IQA algorithms operate without  any reference image information at all.

Popular modern FR picture quality models include the Structural Similarity 
(SSIM)\cite{wang2004image}, its multiscale form,  MS-SSIM\cite{wang2003multiscale}, Visually Information Fidelity (VIF)\cite{sheikh2006image}, FSIM\cite{zhang2011fsim}, GMSD\cite{xue2014gradient},  VSI\cite{zhang2014vsi} and MDSI\cite{nafchi2016mean}. NR-IQA models have also been proposed, including  BRISQUE\cite{mittal2012no},  NIQE\cite{mittal2013making}, BLIINDS\cite{saad2012blind}, and CORNIA\cite{ye2012unsupervised}. 

Several VR-specific IQA models have also been proposed over the years. Yu \textit{et al.}\cite{yu2015framework} proposed a spherical PSNR model called S-PSNR, which averages quality over all viewing directions. The authors of \cite{zakharchenko2016quality} introduced a craster parabolic projection based PSNR (CPP-PSNR) VR-IQA model. Xu \textit{et al.}\cite{xu2018assessing} proposed two kinds of perceptual VQA (P-VQA) methods: a non-content-based PSNR (NCP-PSNR) algorithm and a content-based PSNR (CP-PSNR) method. WS-PSNR\cite{sun2017weighted} is yet another PSNR based VR-IQA method, which reweights pixels according to their location in space. SSIM has also been extended in a similar manner, as exemplified by  S-SSIM\cite{chen2018spherical}. Yang \textit{et al.}\cite{yang2017content} proposed a content-aware algorithm designed specifically to assess stitched VR images, by combining a geometric error metric with a locally-constructed guided IQA method. A NR-IQA method designed to assess stitched panoramic images using convolutional sparse coding and compound feature selection was proposed in \cite{ling2018no}. Given the explosive popularity of deep learning, many more recent VR-IQA methods have been learned to analyze immersive images and videos, often achieving impressive results. For example, in \cite{yang20183d}, the authors deployed an end-to-end 3D convolutional neural network to predict the quality of VR videos without reference. In \cite{lim2018vr} and \cite{kim2019deep}, the power of adversarial learning was utilized to successfully predict the quality of images.
\section{Details of the Subjective Study}
\label{section3}
\subsection{Image Capture}
\begin{figure*} [!ht]
\centerline{
\subfigure[]{
   \label{DMOS}
\includegraphics[width=0.6\columnwidth]{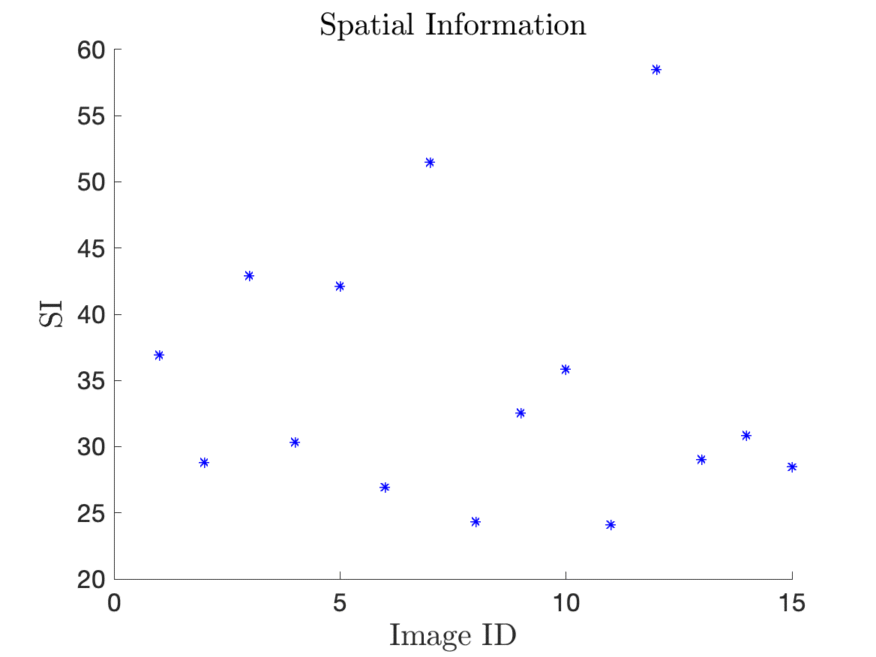}} 
\subfigure[]{
   \label{SROCC}
\includegraphics[width=0.6\columnwidth]{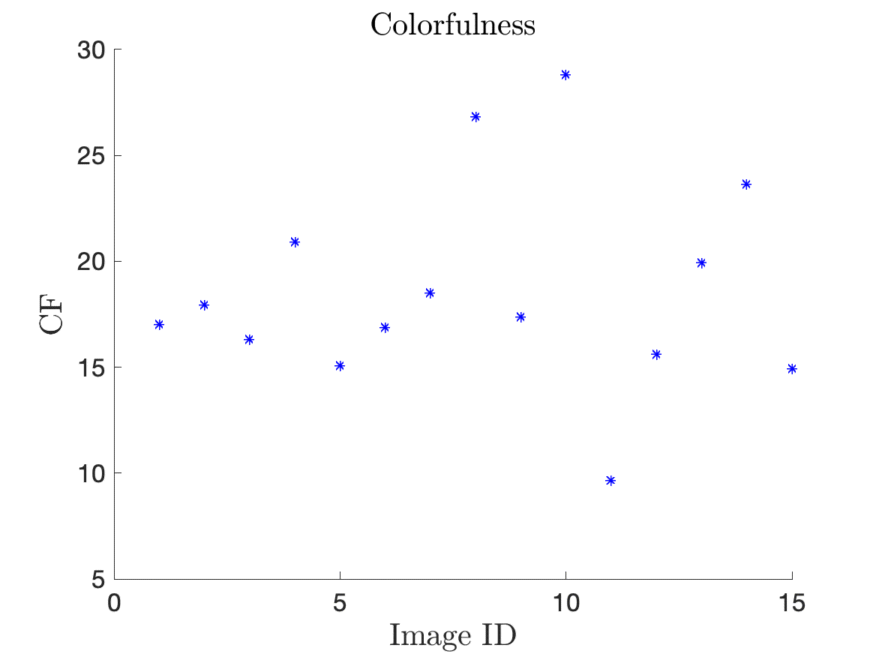}}
\subfigure[]{
   \label{hull}
\includegraphics[width=0.6\columnwidth]{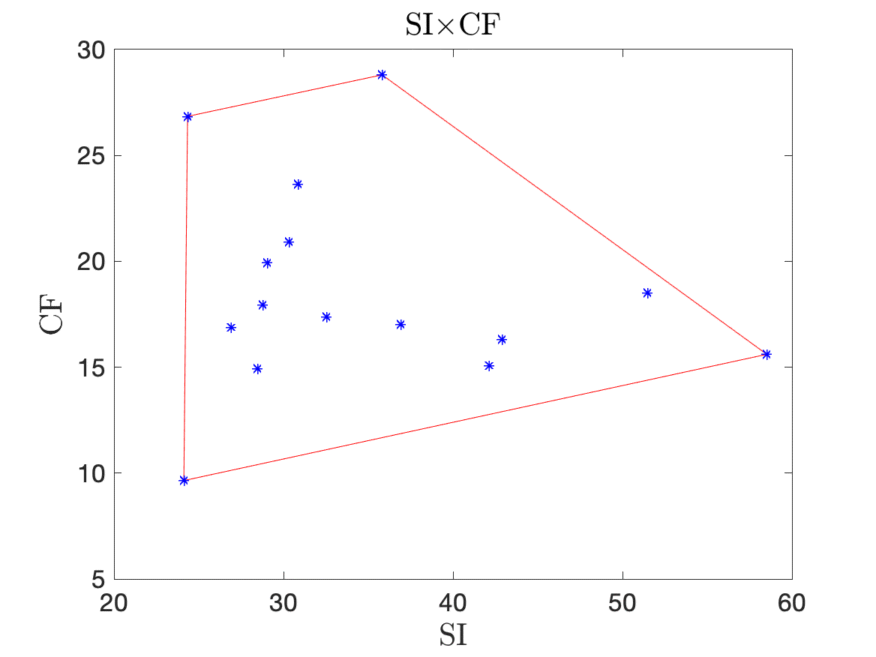}}
}
\caption{Plots of Spatial Information (SI) and Colorfulness (CF) of the VR images in the LIVE VR IQA Database}
\label{fig:si_cf}
\end{figure*}

\begin{figure*} [!ht]
\centerline{
\subfigure[]{
   \label{DMOS}
\includegraphics[width=0.4\columnwidth]{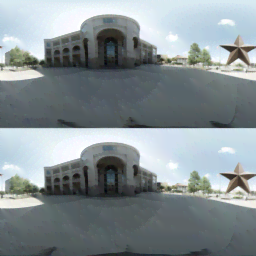}} 
\subfigure[]{
   \label{SROCC}
\includegraphics[width=0.4\columnwidth]{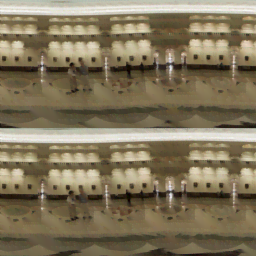}}
\subfigure[]{
   \label{SROCC}
\includegraphics[width=0.4\columnwidth]{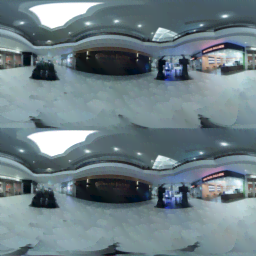}}
\subfigure[]{
   \label{SROCC}
\includegraphics[width=0.4\columnwidth]{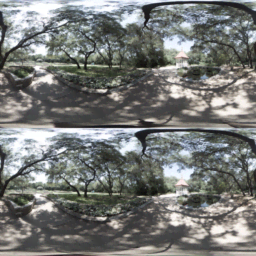}}
\subfigure[]{
   \label{SROCC}
\includegraphics[width=0.4\columnwidth]{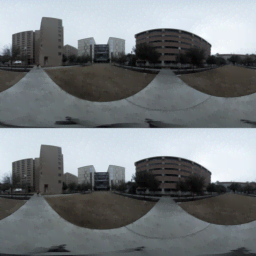}}
}
\centerline{
\subfigure[]{
   \label{DMOS}
\includegraphics[width=0.4\columnwidth]{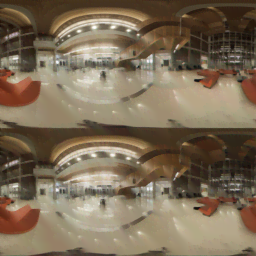}} 
\subfigure[]{
   \label{SROCC}
\includegraphics[width=0.4\columnwidth]{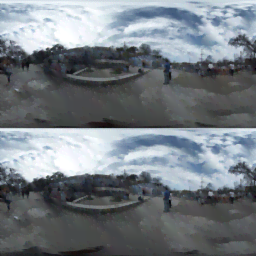}}
\subfigure[]{
   \label{SROCC}
\includegraphics[width=0.4\columnwidth]{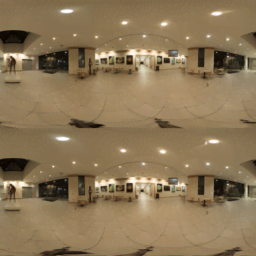}}
\subfigure[]{
   \label{SROCC}
\includegraphics[width=0.4\columnwidth]{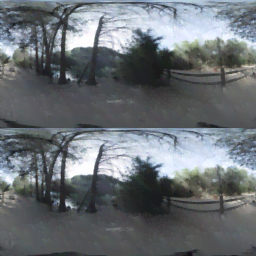}}
\subfigure[]{
   \label{SROCC}
\includegraphics[width=0.4\columnwidth]{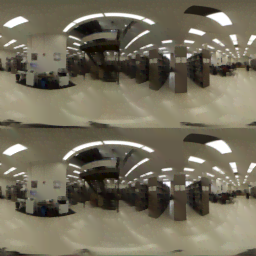}}
}
\centerline{
\subfigure[]{
   \label{DMOS}
\includegraphics[width=0.4\columnwidth]{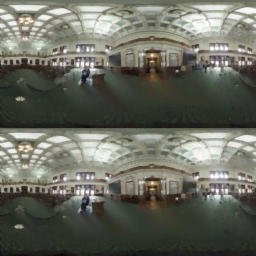}} 
\subfigure[]{
   \label{SROCC}
\includegraphics[width=0.4\columnwidth]{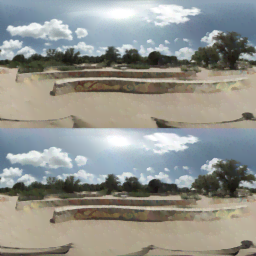}}
\subfigure[]{
   \label{SROCC}
\includegraphics[width=0.4\columnwidth]{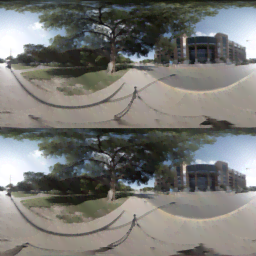}}
\subfigure[]{
   \label{SROCC}
\includegraphics[width=0.4\columnwidth]{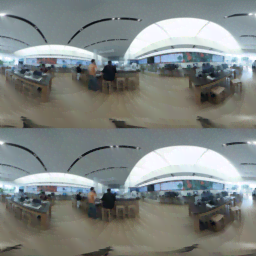}}
\subfigure[]{
   \label{SROCC}
\includegraphics[width=0.4\columnwidth]{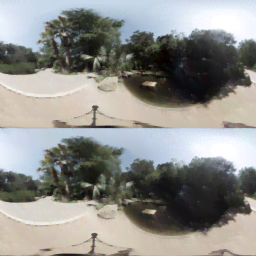}}
}
\caption{Exemplar VR images in the LIVE VR IQA Database}
\label{fig:db_imgs}
\end{figure*}

We used an Insta360 Pro camera \cite{insta360} to capture the VR image in our 360 image database, due to its portability, raw format availability, high resolution (7680 $\times$ 3840), and good image quality. Instead of only capturing colorful, highly saturated images, we collected a wide variety of natural scenes, including daytime/night scenes, sunny/cloudy backgrounds, indoor/outdoor scenes, and so on. We acquired 15 high-quality immersive 3D 360\textdegree{} reference images containing diverse content. Most of the scenes were captured in Austin, Texas. For each scene, 4 to 5 raw images (.dng format) were captured to ensure that one with the least amount of motion blur and stitching error could be selected. For each scene, an over-under equirectangular 3D image was generated. We selected the images to span a wide range of spatial information and colorfulness, as shown in Figure \ref{fig:si_cf}. Spatial Information (SI) is a measure that indicates the amount of spatial detail of a picture, and it is usually higher for more spatially complex scenes\cite{itu1999subjective}. Color information is computed using Colorfulness (CF) as proposed in \cite{hasler2003measuring} which represents intensity and variety of colors in an image. Higher values indicate more colorful images. Figure \ref{hull} depicts a scatter plot of SI vs. CF, showing that our database includes a variety of images considering both metrics. Examples of images in our database are shown in Figure \ref{fig:db_imgs}.  

\begin{figure} [!ht]
\centerline{
\includegraphics[width=0.3\columnwidth]{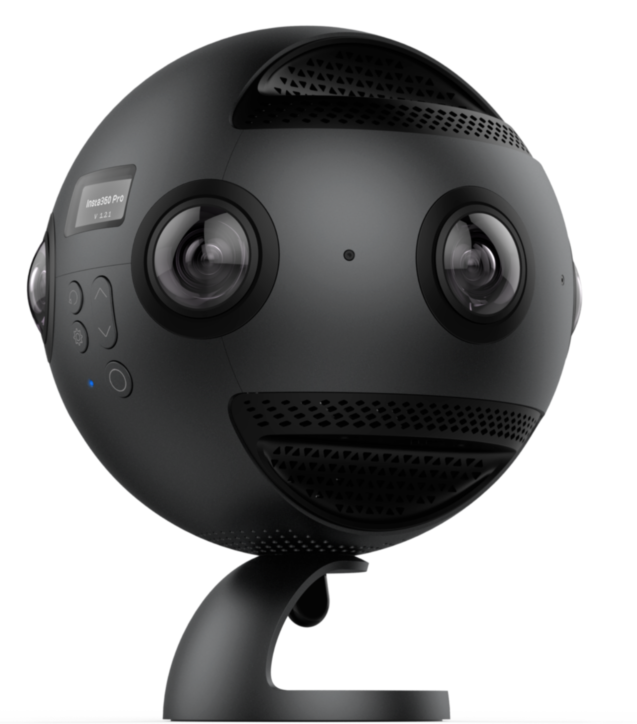}
}
\caption{Insta360 Pro Camera}
\label{fig:insta360}
\end{figure}
\subsection{Test Images}
Each of the selected 15 reference VR content was subjected to 6 types of distortion, including Gaussian noise, Gaussian blur, stitching distortion, downsampling distortion, VP9 compression, and H.265 compression. The driving goal of our study was to create a diverse and representative immersive stereoscopic 3D image quality database for developing, testing, and benchmarking VR-related IQA methods. We included the traditional distortions as well as VR-specific stitching distortions. We also included recent compression distortions, including VP9 and H.265, to study and model the way they compress and perceptually distort VR images. 
 
The distortion levels were determined to ensure noticeable perceptual separation between severity levels while also avoiding obvious differences between neighboring levels. All of the distortions other than stitching distortions were applied directly to the equirectangular 3D image. The 360 images were generated using Insta360 Stitcher. Since the resolution of the original images was 7680 $\times$ 3840, we scaled the reference images to resolution 4096 $\times$ 2048 to match the resolution of the VR headset used in the study (as well as most commercial models) before applying the  distortions. In the following sections, we explain the way each of the different distortions were applied to the 15 reference 3D VR images.
\subsubsection{Gaussian Noise} 
Gaussian additive noise was applied to the unit normalized RGB channels with standard deviations in the range [0.002, 0.03].
\subsubsection{Gaussian Blur}
We separated the left and right images and applied a circular-symmetric 2-D Gaussain kernel to the RGB channels using standard deviations in the range of [0.7, 3.1] pixels. Each RGB channel in both the left and right image was blurred with the same kernel.
\subsubsection{Downsampling}
The left and right images were separated before adding downsampling distortion. Each original immersive image was downsampled to one of five reduced spatial resolutions using bicubic interpolation. We used the HTC Vive for our subjective experiments. This HMD presents a resolution of 1080$\times$1200 and Field of View (FOV) of 110 degrees to each eye. The preferred resolution between 3K and 4K can be found by calculating the portion of solid angle that the FOV spans. We set the maximum total resolution to be 4096$\times$2048, as also suggested in  \cite{guo2018perceptual, zhou2016modeling}, and the minimum resolution to be 820$\times$820, thereby covering a wide range of qualities. 
 
\subsubsection{Stitching Distortion}
\begin{figure} [!ht]
\includegraphics[width=.24\linewidth]{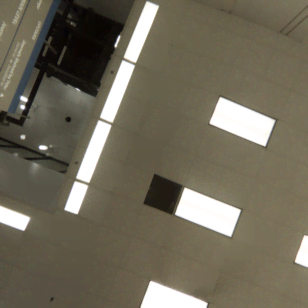} 
\includegraphics[width=.24\linewidth]{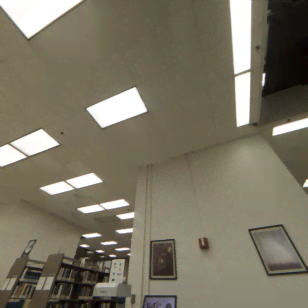}
\includegraphics[width=.24\linewidth]{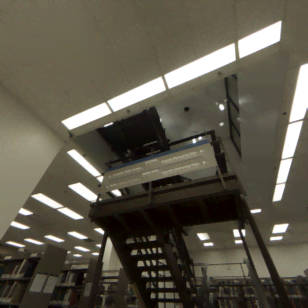}
\includegraphics[width=.24\linewidth]{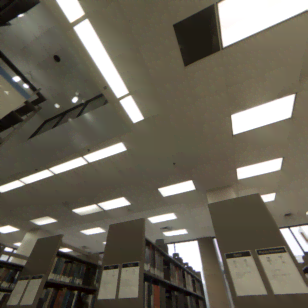}
\centering

\vspace{3.5pt}
\includegraphics[width=.24\linewidth]{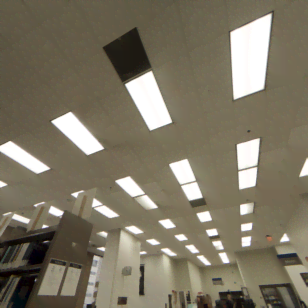}
\includegraphics[width=.24\linewidth]{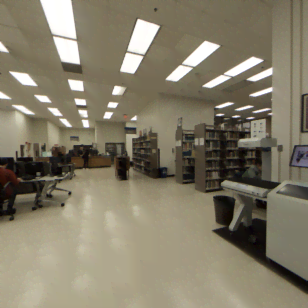}
\includegraphics[width=.24\linewidth]{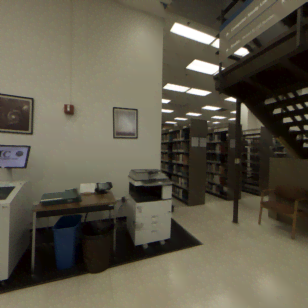}
\includegraphics[width=.24\linewidth]{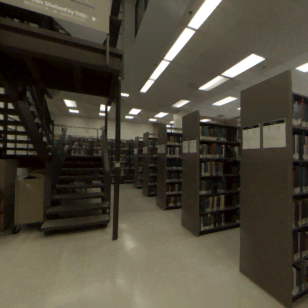}
\centering

\vspace{3.5pt}
\includegraphics[width=.24\linewidth]{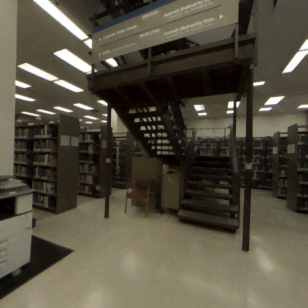}
\includegraphics[width=.24\linewidth]{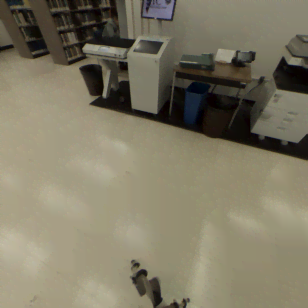}
\includegraphics[width=.24\linewidth]{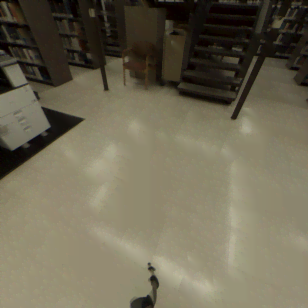}
\includegraphics[width=.24\linewidth]{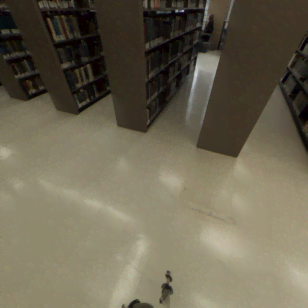}
\centering

\vspace{3.5pt}
\includegraphics[width=.24\linewidth]{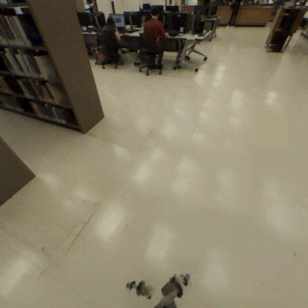}
\includegraphics[width=.24\linewidth]{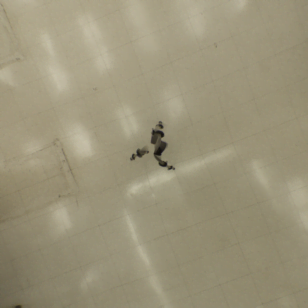}
\caption{Example of 14 perspective views that were stitched together}
\label{fig:perspective}
\end{figure}

We first separated the left and right images and captured 14 perspective views from each image using MATLAB, covering the entire spherical image to simulate a 14-head panoramic camera placed at the center of each scene \cite{yang2017content}. The viewing directions we used are listed in Table \ref{directions}, where $\phi$ represents the zenith angle, and $\theta$ represents the azimuth angle. The FOV was set to 110 degrees. An example of the 14  views is shown in
Figure \ref{fig:perspective}.

After obtaining a set of images captured by the virtual lenses, we imported the views into the popular stitching tool Nuke, and adjusted the orientation of each stitched image to have the same orientation as its reference image, to avoid introducing any further discomfort. Specifically, since the first viewing direction points to the zenith, we adjusted the ZXY rotation
parameters in Nuke such that the first perspective view (generated by the first viewing direction) was on the right position. This was done by searching the rotation matrix space to  find the parameters that would rotate the first view back to the zenith. 

\begin{figure*} [!ht]
\centerline{
\subfigure[]{
   \label{}
   \includegraphics[width=0.65\columnwidth]{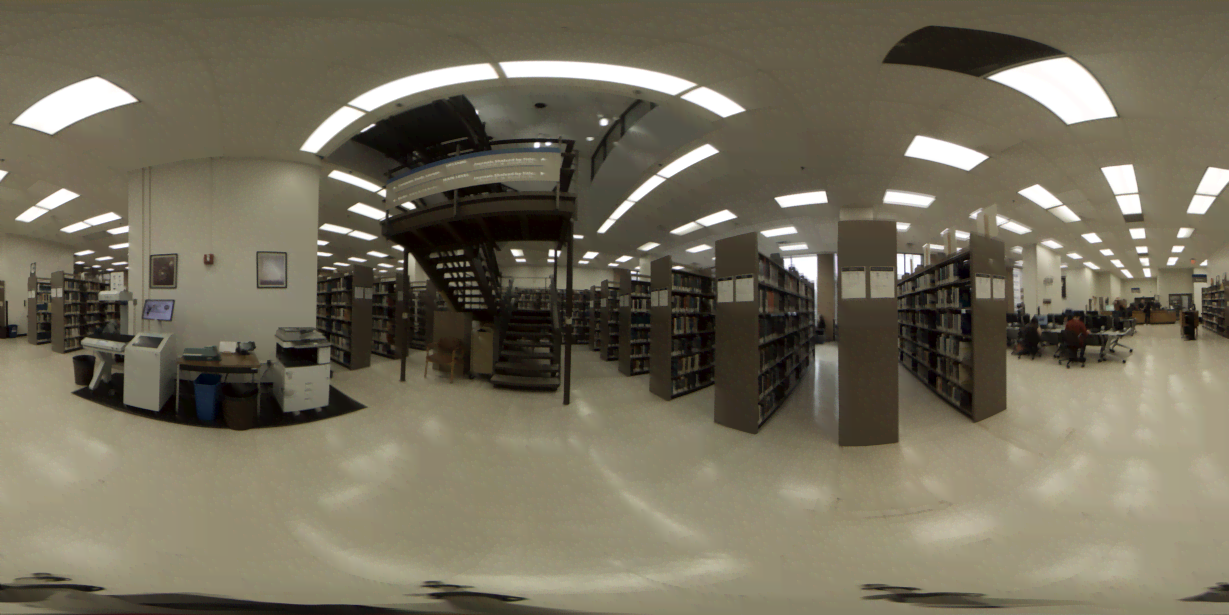}}
\subfigure[]{
   \label{}
   \includegraphics[width=0.65\columnwidth]{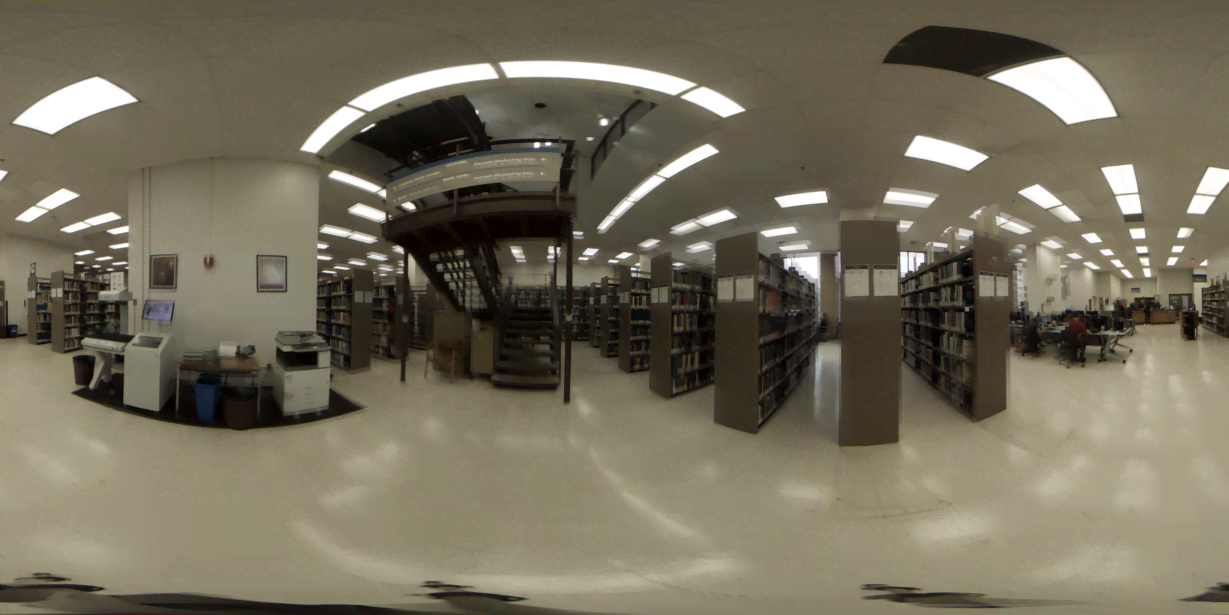}}
\subfigure[]{
   \label{}
   \includegraphics[width=0.65\columnwidth]{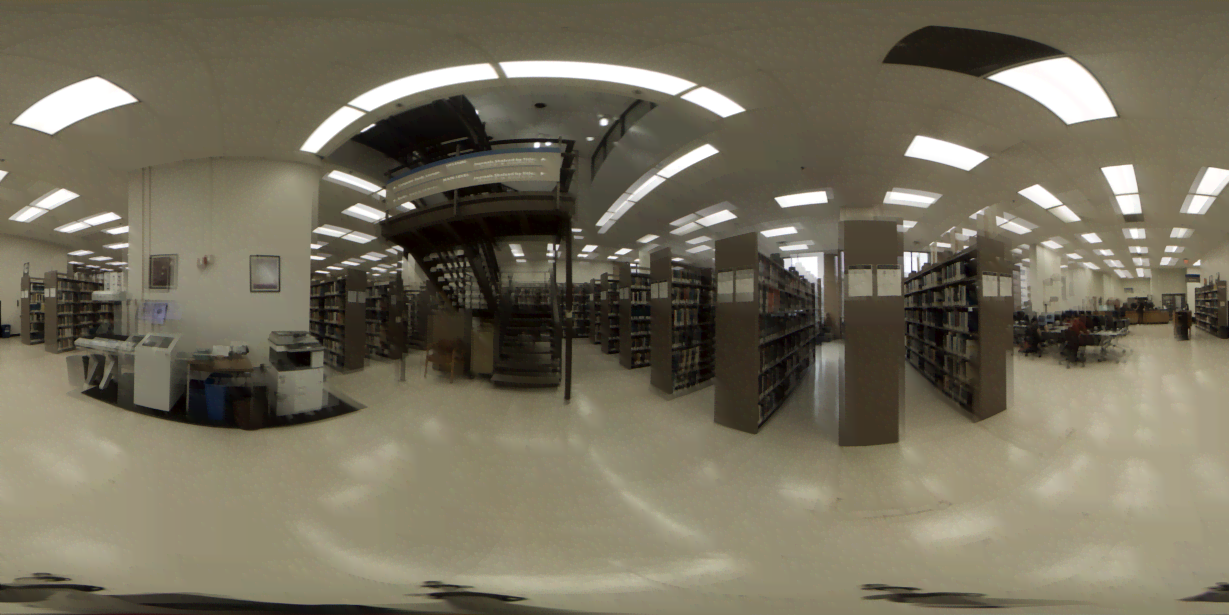}}
   }
\centering
\subfigure[]{
   \label{}
   \includegraphics[width=0.65\columnwidth]{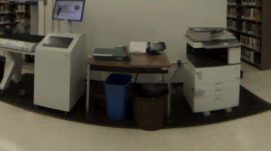}}
\subfigure[]{
   \label{}
   \includegraphics[width=0.65\columnwidth]{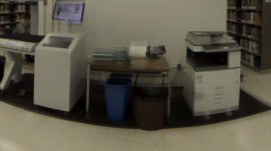}}
\subfigure[]{
   \label{}
   \includegraphics[width=0.65\columnwidth]{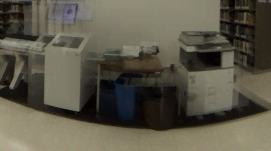}}
\caption{Different levels of stitching distortion. (a)-(c): Images of level 1, 3 and 5 (higher levels indicate more distortion). (d)-(e): Zoomed-in views of (a)-(c).}
\label{fig:st}
\end{figure*}

After adjusting the orientation of the stitched image, we tuned the stitching parameters, mainly the convergence
distance, error threshold and whether `refine' or  `reject' was applied, to generate different
levels of the distortion. An example of different levels of stitching distortion is shown in Figure \ref{fig:st}. The same procedure was applied on the left and right images, and we ensured that the stitching distortion created was at the same location in the two images to avoid further discomfort arising from binocular rivalry.
\begin{table*}[htp]
\centering
\caption{Viewing Directions, where $\phi$ represents the zenith angle, and $\theta$ represents the azimuth angle}
\label{directions}
\begin{tabular}{|c|c|c|c|c|c|c|c|c|c|c|c|c|c|c|}
\hline
        $\theta$ & 0  & $\pi/4$ & $3\pi/4$& $5\pi/4$& $7\pi/4$& 0& $\pi/2$& $\pi$& $3\pi/4$& $\pi/4$& $3\pi/4$& $5\pi/4$& $7\pi/4$& $0$   \\ \hline
$\phi$ & 0  & $\pi/4$ & $\pi/4$& $\pi/4$& $\pi/4$&  $\pi/2$& $\pi/2$& $\pi/2$& $\pi/2$& $3\pi/4$& $3\pi/4$& $3\pi/4$& $3\pi/4$&$\pi$ \\ \hline
\end{tabular}
\end{table*}

\subsubsection{VP9 Compression}
VP9 compression was applied using the popular public domain software FFmpeg, using the libvpx-vp9 encoder. We varied the constant quality factor over the range [50, 63], where lower values indicate better quality.
\subsubsection{H.265 Compression}
H.265 (HEVC) compression distortion was applied using the FFmpeg libx265 encoder with different QP values ranging from 38 to 50, where higher values imply increased compression and worse quality. 
\subsection{Subjective Testing Design}
We employed the Single Stimulus Continuous Quality evaluation methods described in the ITU-R BT 500.13 recommendation\cite{ITU}. The human subjects entered their quality adjustments on a  continuous rating scale from 0 to 100, where 0 indicates worst quality. 
 
Each viewing session was limited to a duration of 30 minutes and the subjects were free to take rests at any time. The subjects were asked whether they were prone to discomfort when participating in either a VR or 3D environment beforehand, to eliminate subjects who were not suitable for this subjective study. The visual acuity of each subject was determined using the Snellen test, and each subject was asked to wear their corrective lenses to achieve normal vision when participating in the study. Each subject also participated in a RanDot Stereo test of their stereo vision and depth perception. If any test showed impairment, the subject was recommended not to take this test, but if the subject decided to perform the test, the results were discarded. The range of Interpupillary Distances (IPD) of the HTC Vive is 60.3mm-73.7mm. For those subjects whose IPD was outside of this range, a period of experimentation with the HMD was allowed. If the subject felt uncomfortable, then it was recommended that he/she not perform the test. The data was also discarded when the subject did not follow instructions. 

Each subject participated in three sessions separated by at least 24 hours apart. For each session, 9 contents and 60 distorted images were randomly selected. The "hidden" reference image was included in each session. To reduce the effects of memory comparisons, images of the same content were separated by at least five images of different content. The average viewing time for each session was 27 minutes, with the average viewing and rating time for each image being around 23 seconds.

\subsection{Subjective Test Display}
\begin{figure} [!ht]
\centerline{
\includegraphics[width=0.5\columnwidth]{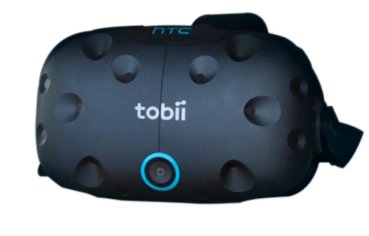}
}
\caption{HTC Vive integrated with the Tobii Pro Eye Tracking system.}
\label{fig:vive}
\end{figure}

The subjective test was displayed on a HTC Vive VR headset with a built-in Tobii Pro eye tracking system\cite{tobiicalibration}, as depicted in Figure \ref{fig:vive}. The Tobii Pro Eye tracking is fully integrated into the HTC Vive HMD. It trackes the gaze direction using the Pupil Center Corneal Reflection technique. More specifically, it uses dark pupil eye tracking, where an illuminator is placed away from the optical axis causing the pupil to appear darker than the iris. Tobii Pro eye tracking has an accuracy of 0.5\degree{}, a latency of approximately 10ms, and a sampling frequency of 120 Hz. There are several data outputs for each eye: device and system timestamp, gaze origin, gaze direction, pupil position and absolute pupil size. Image playback was supported by a dedicated high performance server (Intel i7-6700, 32GB memory, 1TB hard drive, NVIDIA TITAN X). The interface was built using Unity Game Engine. Detailed procedures of the subjective test are described in the following sections.
\subsubsection{Eye tracking}
\begin{figure} [!ht]
\centerline{
\includegraphics[width=0.5\columnwidth]{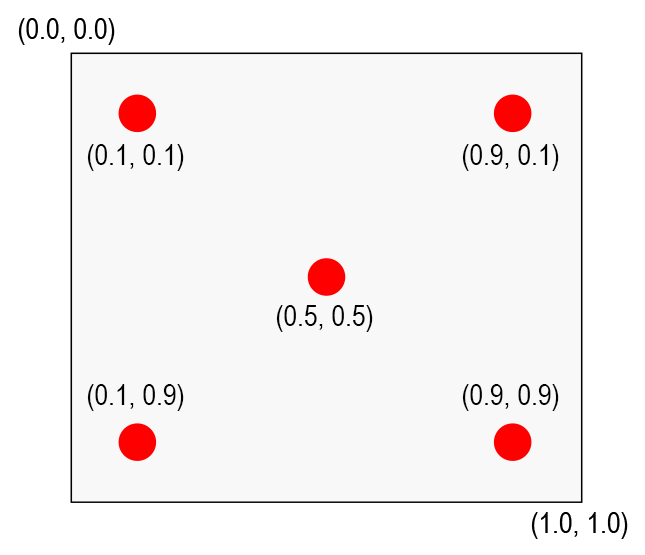}
}
\caption{Calibration pattern.}
\label{fig:cali}
\end{figure}

Eye tracking commenced at the beginning of each session. Subjects fixated on five red dots that flashed sequentially in the HMD at different positions\cite{tobiicalibration}, as shown in Figure \ref{fig:cali}. These points are mapped in normalized coordinates so that (0.0, 0.0) corresponds to the upper left corner and (1.0, 1.0) corresponds to the lower right corner of the current viewport. Each subject was asked to stare at each dot in succession, then after the last dot disappeared, the system used the recorded dot fixations to calibrate the eyetracker. The process was repeated if the calibration was not successful. If the calibration was still not successful after five trials, the subject would be asked to participate at another time. This situation happened twice during our experiments.
\subsubsection{Viewing and Scoring}
\begin{figure} [!ht]
\centerline{
\includegraphics[width=0.7\columnwidth]{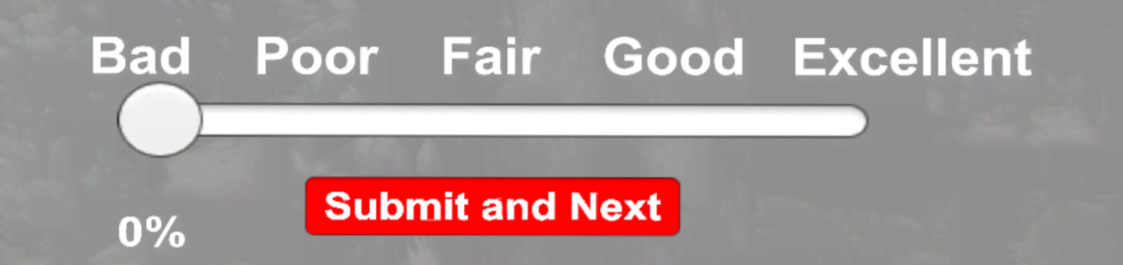}
}
\caption{Rating bar used in the subjective study}
\label{fig:GUI}
\end{figure}

The quality scale popped up automatically after 20 seconds of viewing to limit each subject's time viewing the images. To avoid having the subject view the image after the time limit, a grey canvas displayed as background of the rating bar, as shown in Figure \ref{fig:GUI}. The quality scale was in the center of the subject's field of view, wherever they moved their head. Five Likert labels ``Bad, Poor, Fair, Good, Excellent" indicated the range of ratings the subject could apply. To rate the images, the subjects used the hand controllers supplied with the VR headset to choose the desired score on the quality scale. After the subject was satisfied with the score chosen, they clicked on 'Submit and Next' to see the next image. Once the subject submitted the score, the name and score of the image were written to file. The submission timestamp was also recorded to determine the correspondences between the gaze data and the image. The subsequent image was randomly chosen from all the images in the session, subject to the previously mentioned constraints on the display order. Detailed gaze data was output by the Tobii Pro at the end of each session. 

\subsection{Subjects and Training}
All subjects were students at The University of Texas at Austin. The subject pool was inexperienced with image quality assessment and image distortions. A total of 40 students were involved in the study, and each image was rated by around 15 students.

Each subject was orally briefed about the goals of the study and presented with the detailed procedure in written form. A consent form was also signed by the subject. Each subject was asked to view the image as much as possible and score the images according to image quality only, without regard to the appeal of the content. Before the actual session, each subject viewed a training session of 10 images not included in the database. These images were distorted in the same way as the images in the database and spanned the same ranges of quality, to give the subject an idea of the quality and distortions that would be seen in the actual sessions. The subjects rated these images accordingly using the same technique as in the actual session to familiarize themselves with the controllers and the VR headset.  

\section{Data Analysis}
\label{section4}
Subjective Difference Mean Opinion Score (DMOS) were computed according to \cite{seshadrinathan2010study}.The difference scores for reference images were 0 and were discarded for all sessions. Then  per session Z-scores were computed from the difference scores and combined into a score matrix {$z_{ij}$} and a "viewed" matrix {$s_{ij}$}, where 0 indicates the image was not seen by the subject and 1 indicates the image was seen by the subject.

Subject rejection was performed using the ITU-R BT 500.11\cite{ITU} to discard unreliable subjects. To proceed with subject rejection, we first determined whether the scores assigned by a subject were normally distributed, using the $\beta_{2}$ test by calculating the kurtosis coefficient of the function:
\begin{equation*}
    \beta_{2,j} = \frac{m_4}{(m_2)^2}
\end{equation*}
and 
\begin{equation*}
    m_{x} = \frac{\sum_{i = 1}^{M_{view}}{(z_{j} - \Bar{z_{j}})^x}}{M_{view}},
\end{equation*}
where $M_{view}$ is the number of subjects that have seen image $j$. 
We calculated the mean score and standard deviation for each image:
\begin{equation*}
    \Bar{z_{j}} = \frac{1}{M_{view}}\sum_{i = 1}^{M_{view}}{z_{ij}}
\end{equation*}
\begin{equation*}
    \sigma_{j} = \sqrt{\sum_{i = 1}^{M_{view}}\frac{(z_{j} - \Bar{z_{j}})^2}{M_{view} - 1}}
\end{equation*}
If $\beta_{2}$ fell between 2 and 4, the scores were assumed to be normally distributed. Then:

if $z_{j} \geq \Bar{z_{j}} + 2\sigma_{j}$, then $P_i = P_i + 1$

if $z_{j} \leq \Bar{z_{j}} - 2\sigma_{j}$, then $Q_i = Q_i + 1$

If the scores were deemed to not be normally distributed, then:

if $z_{j} \geq \Bar{z_{j}} + \sqrt{20}\sigma_{j}$, then $P_i = P_i + 1$

if $z_{j} \leq \Bar{z_{j}} - \sqrt{20}\sigma_{j}$, then $Q_i = Q_i + 1$

To reject a subject, we determined whether the following two conditions hold:
\begin{equation}
    \frac{P_i + Q_i}{N} > 0.5,
\label{reject_1}
\end{equation}
where $N$ is the number of images in the study,
and 
\begin{equation}
    \left|\frac{P_i - Q_i}{P_i + Q_i}\right| < 0.3.
\label{reject_2}
\end{equation}
If Equation \ref{reject_1} and Equation \ref{reject_2} were both found to hold, then a subject was rejected.

\begin{figure*} [!ht]
\centerline{
\subfigure[]{
   \label{DMOS}
\includegraphics[width=0.8\columnwidth]{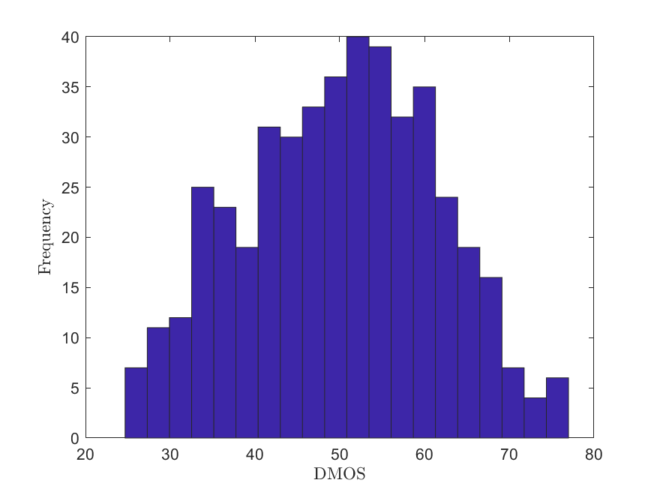}} 
\subfigure[]{
   \label{SROCC}
\includegraphics[width=0.8\columnwidth]{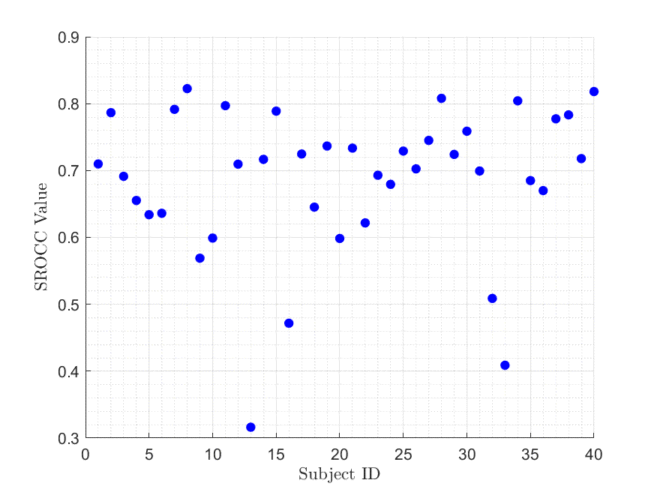}}
}
\caption{(a) Histogram of DMOS. (b) SROCC between subject ratings and DMOS.}
\label{fig:DMOS_SORCC}
\end{figure*}

In our study, 2 out of 42 subjects were rejected. For the remaining subjects, we mapped their Z-score to [0, 100] using equation mentioned in \cite{seshadrinathan2010study}. Finally, the DMOS of each image was obtained by computing the mean of the rescaled Z-scores from 40 remaining subjects.
A histogram of the recorded DMOS and a plot of the correlations between each subject's ratings and DMOS are shown in Figure \ref{fig:DMOS_SORCC}. The DMOS were found to lie in the range [24.67, 76.99].

To explore the internal consistency of the subject data, we randomly divided the subjects into two equal size groups, and computed the Spearman's Rank Correlation Coefficient (SROCC) correlation between their scores. This was done 1000 times. After 1000 splits, the range of correlations was found to be between 0.80 and 0.90 with a median  value of 0.87. Hence, there was a high degree of inter-subject agreement despite the more complex immersive viewing environment. We also calculated correlations by distortion category as shown in Table \ref{correlation}. Clearly, stitching distortion resulted in the lowest inter-subject correlation, which is not unexpected, since stitching distortions are highly localized distortions and their ratings are dependent on the amount of visual attention they received from each subject.

\begin{table*}[htp]
\centering
\caption{Min, Max and Median SROCC between randomized subject groups for each distortion category}
\label{correlation}
\begin{tabular}{|c|c|c|c|c|c|c|}\hline
& GAUSSIAN BLUR & GAUSSIAN NOISE & DOWNSAMPLING & STITCHING & VP9 & H.265 \\ \hline
MIN & 0.7778 & 0.6492 & {\textbf{0.8640}} & {\textbf{0.5669}}  & 0.6056 & 0.8173 \\\hline
MAX & 0.9316 & 0.8815 & {\textbf{0.9564}} & {\textbf{0.8535}} & 0.8793 & 0.9432 \\\hline
MEDIAN & 0.8625 & 0.7897 & {\textbf{0.9146}} & {\textbf{0.7184}} & 0.7746 &  0.8951\\\hline
\end{tabular}
\end{table*}

\begin{figure*} [htp]
\centering
\subfigure[Gaussian Blur]{
   \label{trend_gb}
\includegraphics[width=0.6\columnwidth]{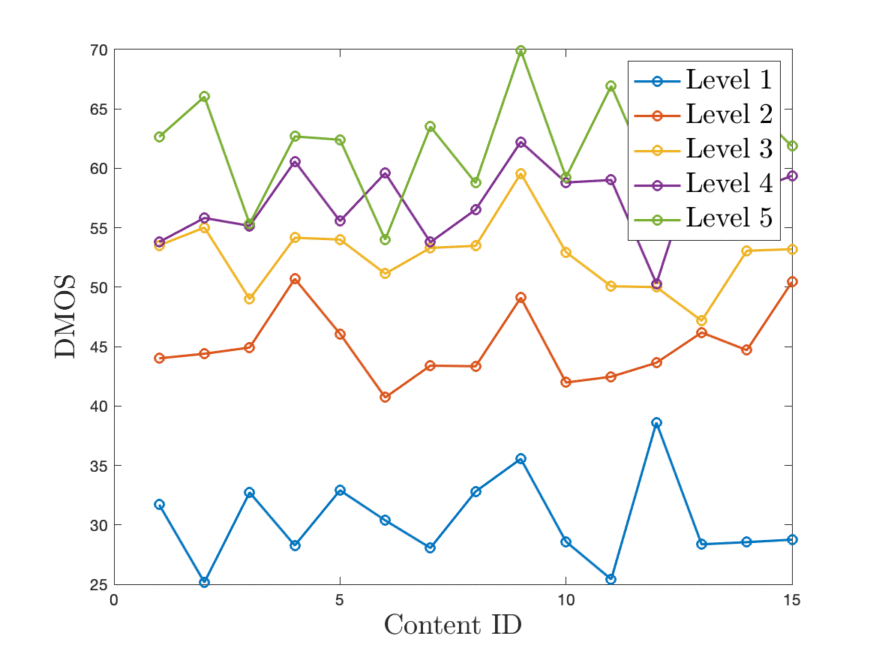}} 
\subfigure[Gaussian Noise]{
   \label{trend_gn}
\includegraphics[width=0.6\columnwidth]{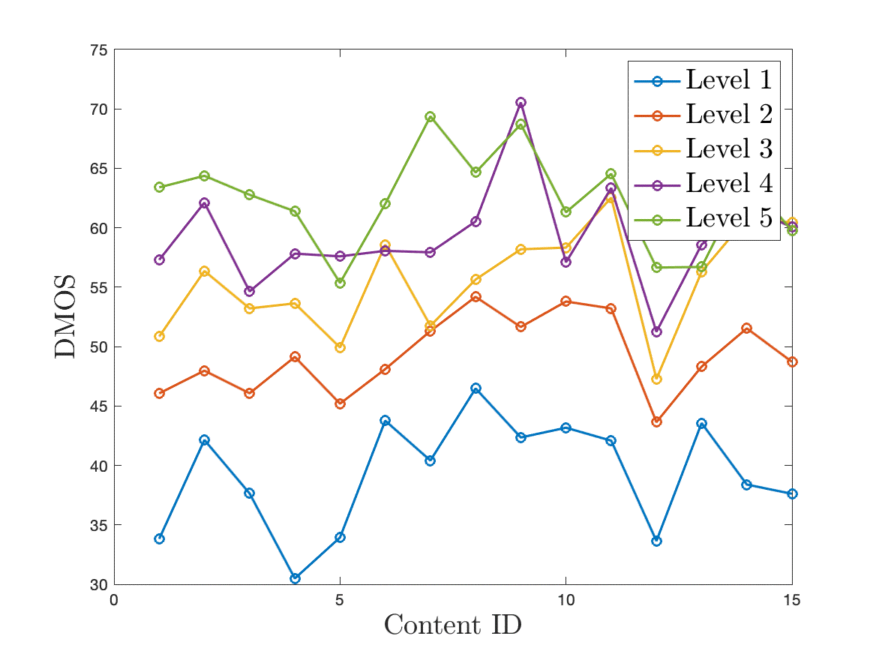}} 
\subfigure[Downsampling]{
   \label{trend_ds}
\includegraphics[width=0.6\columnwidth]{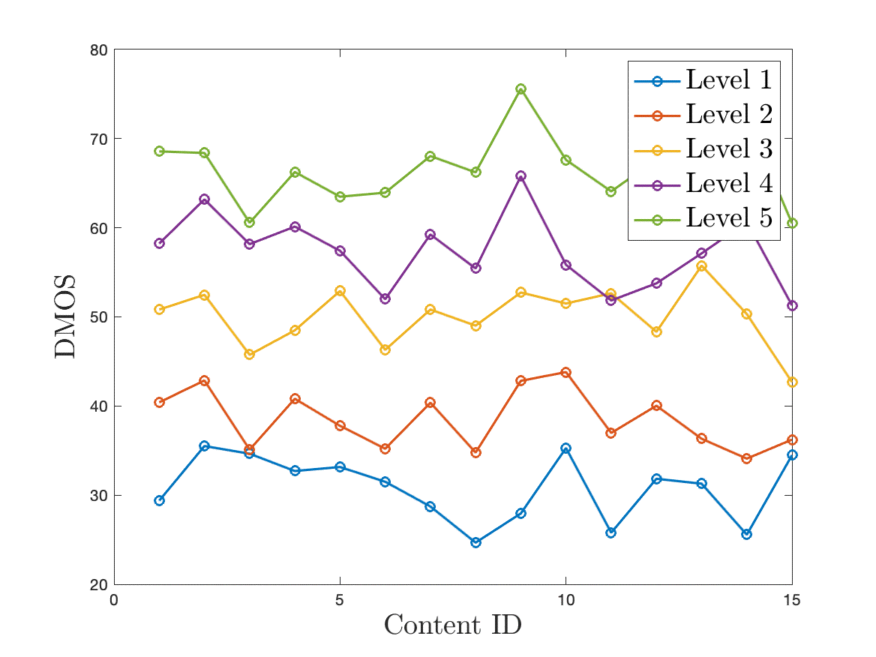}} 
\subfigure[Stitching Distortion]{
   \label{trend_st}
\includegraphics[width=0.6\columnwidth]{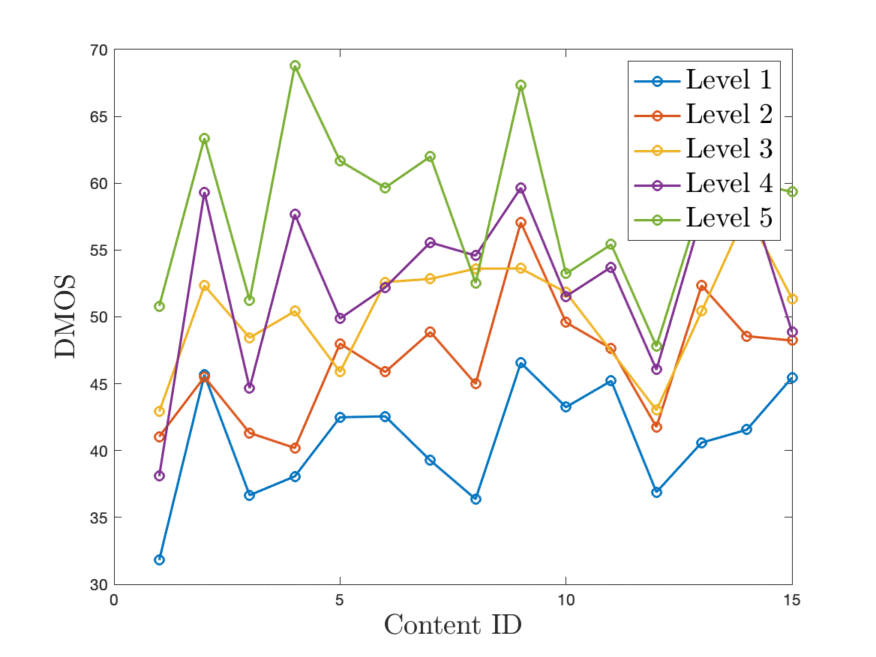}} 
\subfigure[VP9]{
   \label{trend_vp9}
\includegraphics[width=0.6\columnwidth]{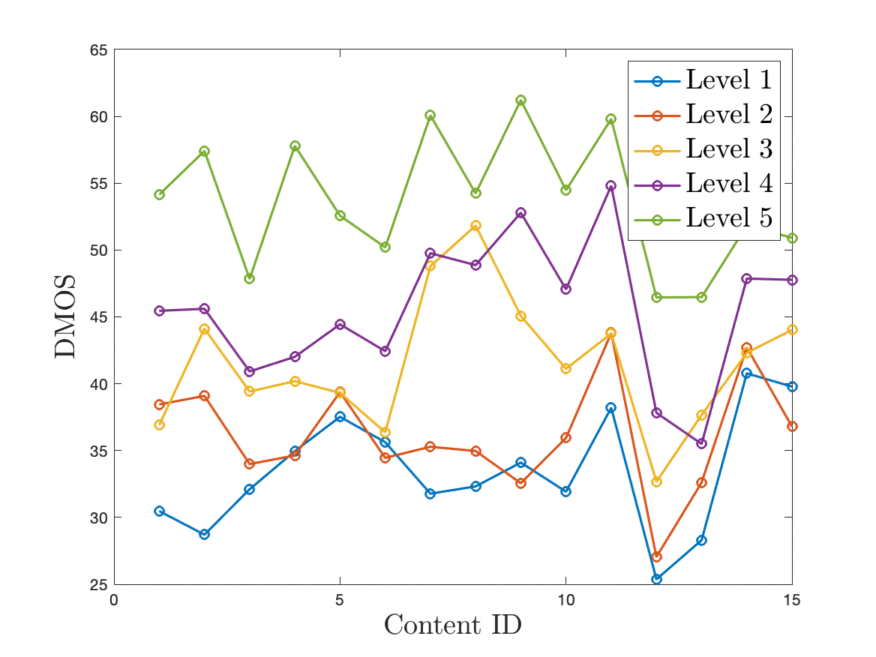}} 
\subfigure[H.265]{
   \label{trend_x265}
\includegraphics[width=0.6\columnwidth]{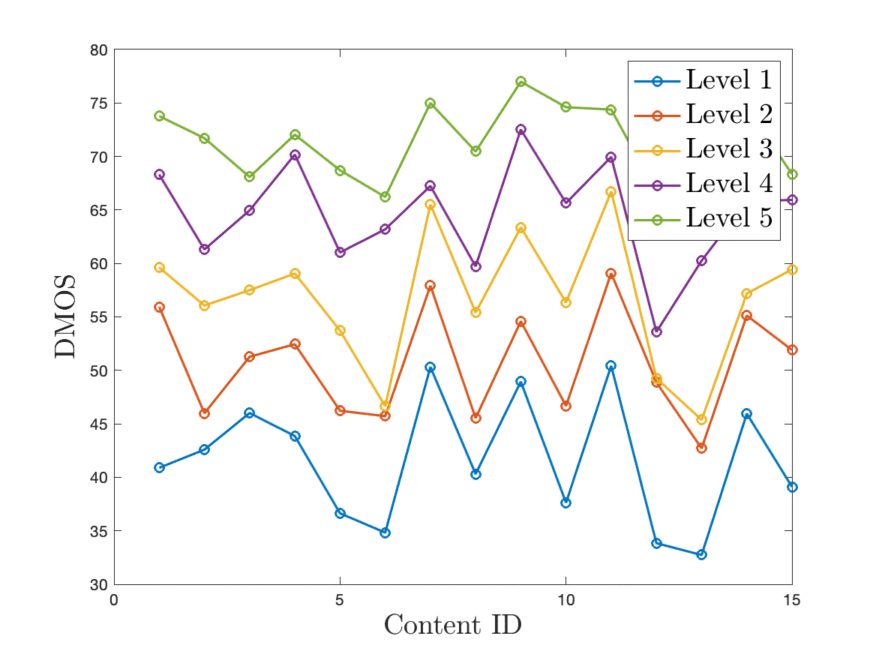}} 
\caption{DMOS of all contents for each level of applied distortion.}
\label{fig:DMOS_Trend}
\end{figure*}
Figure \ref{fig:DMOS_Trend} plots the DMOS  across all contents, where each color coded curve corresponds to a different distortion level. As shown in the figure, for downsampling and H.265 compression distortions, the DMOS  associated with most of the contents decreased with distortion level and the DMOS for different distortion levels are clearly separated. Interestingly, for Gaussian noise and VP9 distortions, the DMOS given to some of the contents were not always monotonic with distortion level. For stitching distortions, the DMOS across distortion levels were mostly consistent but slighly entangled.

\begin{figure*} [htp]
\centering
\subfigure[Gaussian Blur]{
   \label{errorbar_gb}
\includegraphics[width=0.6\columnwidth]{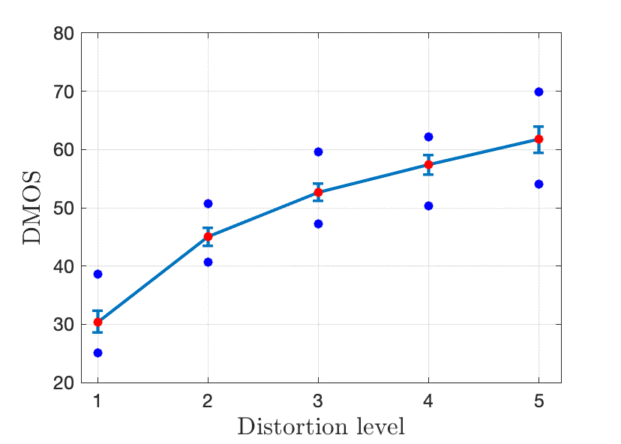}} 
\subfigure[Gaussian Noise]{
   \label{errorbar_gn}
\includegraphics[width=0.6\columnwidth]{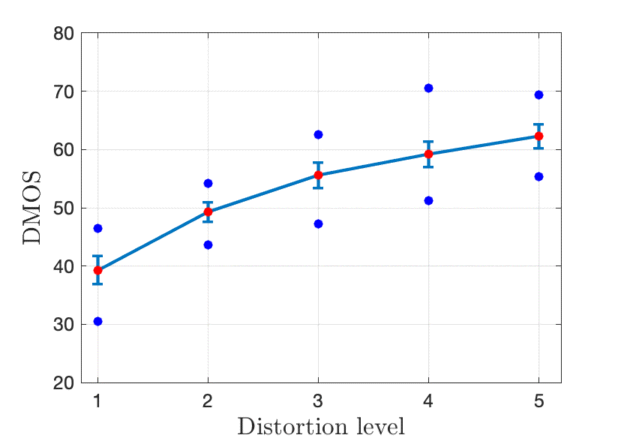}} 
\subfigure[Downsampling]{
   \label{errorbar_ds}
\includegraphics[width=0.6\columnwidth]{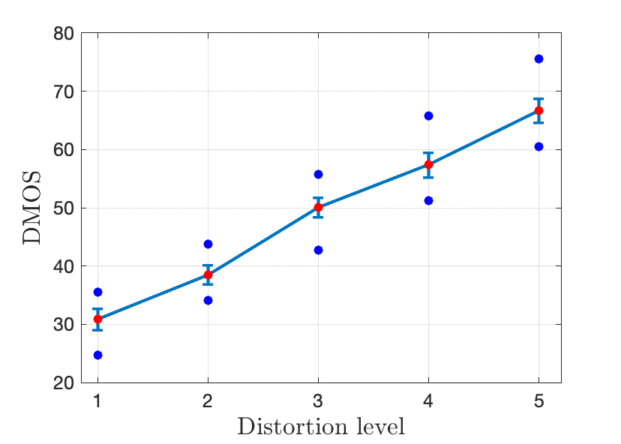}} 
\subfigure[Stitching Distortion]{
   \label{errorbar_st}
\includegraphics[width=0.6\columnwidth]{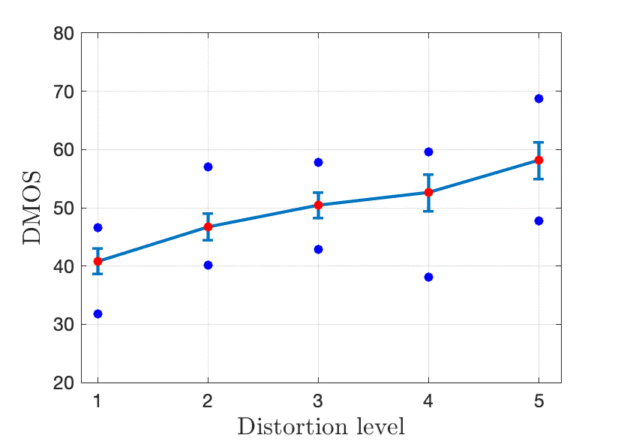}} 
\subfigure[VP9]{
   \label{errorbar_vp9}
\includegraphics[width=0.6\columnwidth]{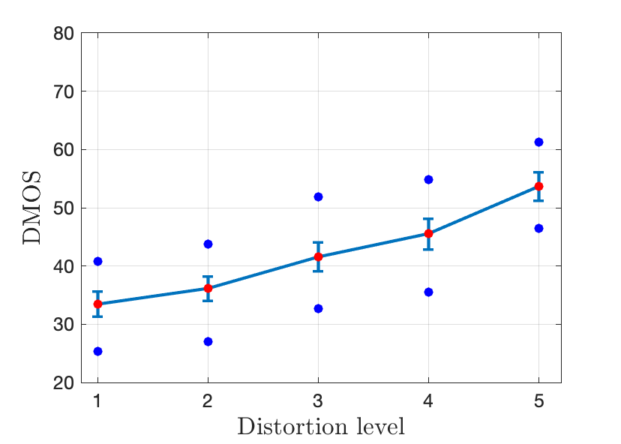}} 
\subfigure[H.265]{
   \label{errorbar_x265}
\includegraphics[width=0.6\columnwidth]{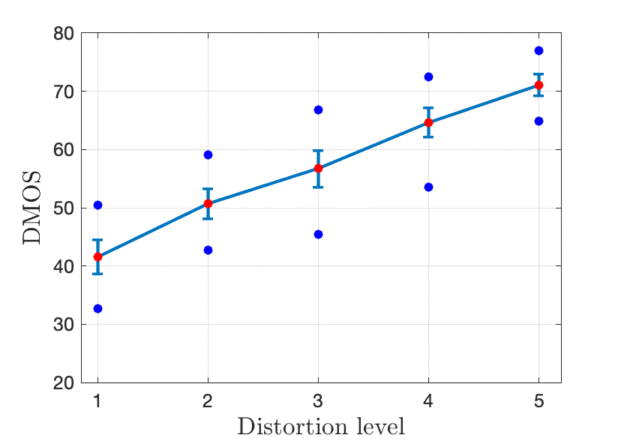}} 
\caption{Confidence intervals of DMOS over all contents for each applied level of distortion. The blue points indicate the maximum and the minimum DMOS for each distortion type and level. The red points indicate the mean DMOS and the blue bars are the 95\% confidence intervals.}
\label{fig:errorbar}
\end{figure*}

Figure \ref{fig:errorbar} plots the DMOS ranges against distortion level for each distortion type. There were overlaps of the confidence intervals for Gaussian noise, stitching and VP9 distortions. Overlaps occurred at higher distortion levels for Gaussian noise, at lower distortion levels for VP9 and over all regions for stitching distortions. This indicates that more severe Gaussian noise distortions as were light VP9 distortions were rated similarly, while stitching distortions were less consistently rated overall. 

\begin{figure*} [htp]
\centering
\subfigure[PSNR]{
   \label{errorbar_gb}
\includegraphics[width=0.6\columnwidth]{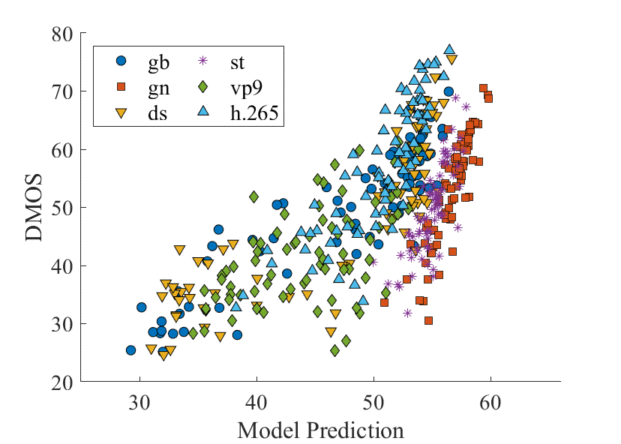}} 
\subfigure[WS-PSNR]{
   \label{errorbar_gn}
\includegraphics[width=0.6\columnwidth]{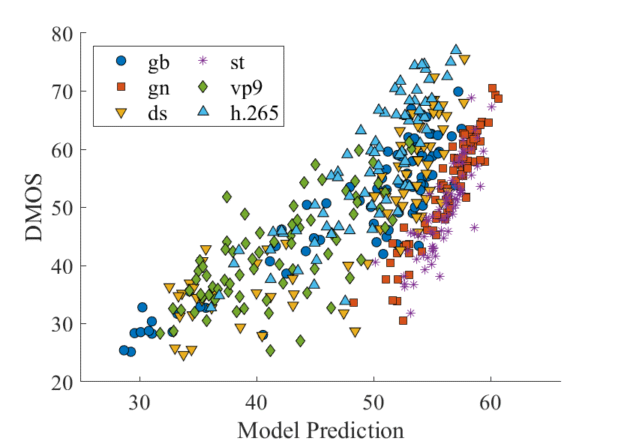}} 
\subfigure[SSIM]{
   \label{errorbar_ds}
\includegraphics[width=0.6\columnwidth]{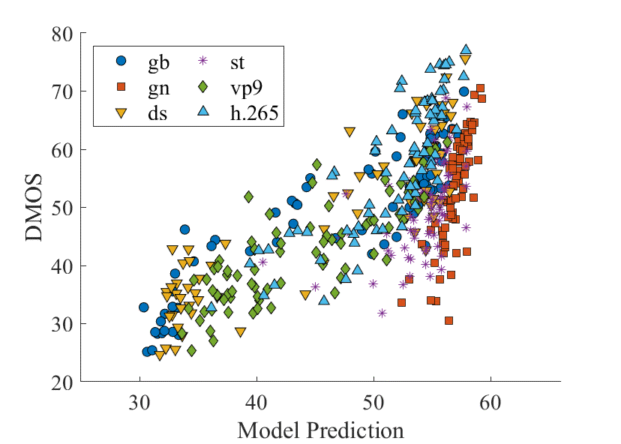}} 
\subfigure[MS-SSIM]{
   \label{errorbar_st}
\includegraphics[width=0.6\columnwidth]{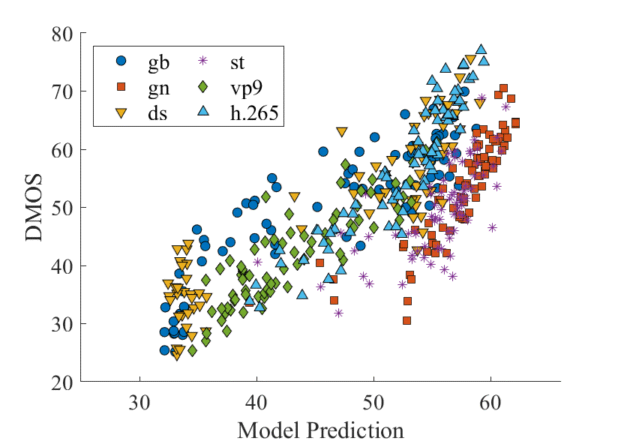}} 
\subfigure[S-SSIM]{
   \label{errorbar_vp9}
\includegraphics[width=0.6\columnwidth]{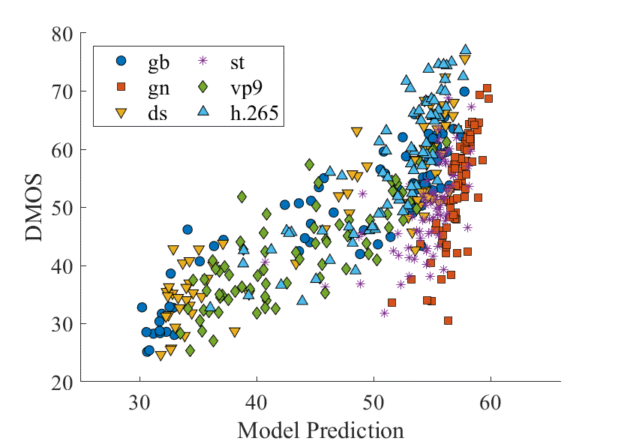}} 
\subfigure[FSIM]{
   \label{errorbar_x265}
\includegraphics[width=0.6\columnwidth]{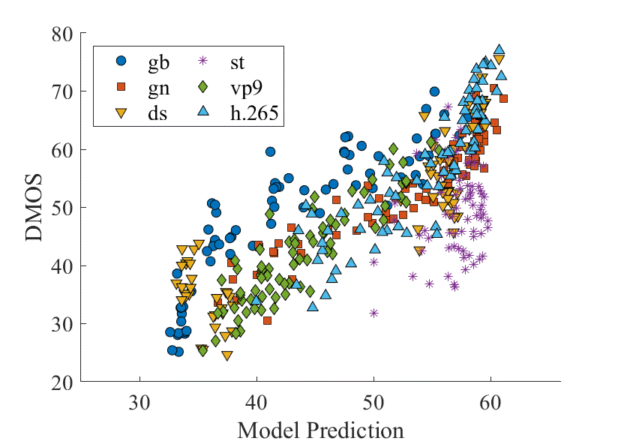}} 
\subfigure[VSI]{
   \label{errorbar_x265}
\includegraphics[width=0.6\columnwidth]{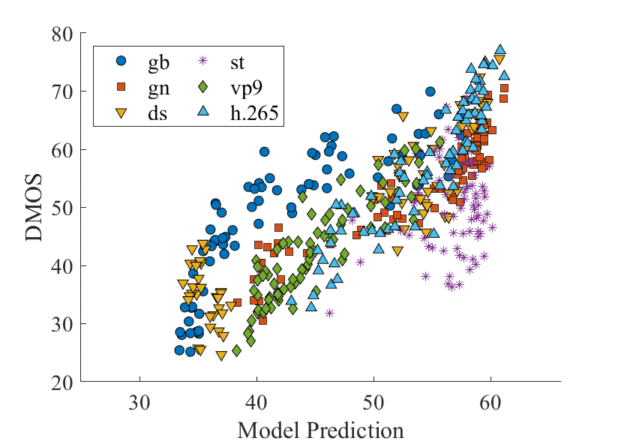}} 
\subfigure[GMSD]{
   \label{errorbar_x265}
\includegraphics[width=0.6\columnwidth]{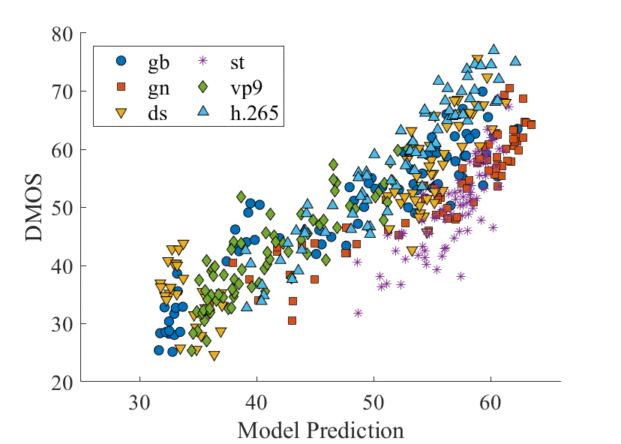}}
\subfigure[MDSI]{
   \label{errorbar_x265}
\includegraphics[width=0.6\columnwidth]{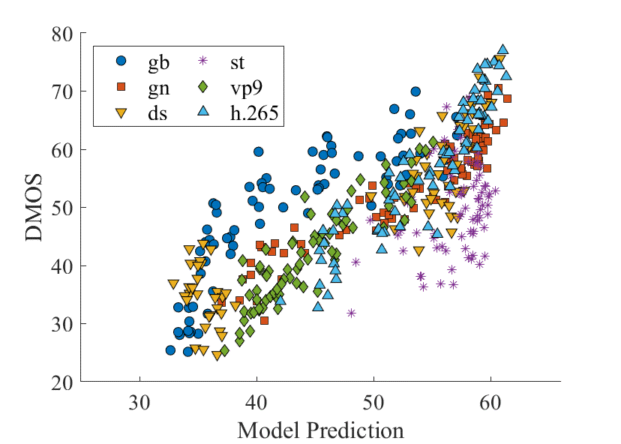}}
\subfigure[BRISQUE]{
   \label{errorbar_x265}
\includegraphics[width=0.6\columnwidth]{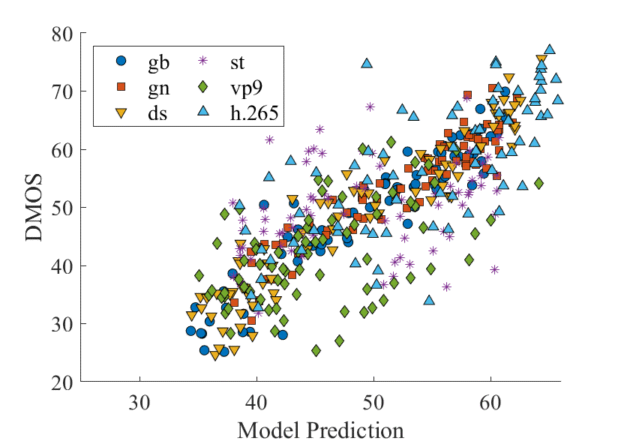}}
\subfigure[NIQE]{
   \label{errorbar_x265}
\includegraphics[width=0.6\columnwidth]{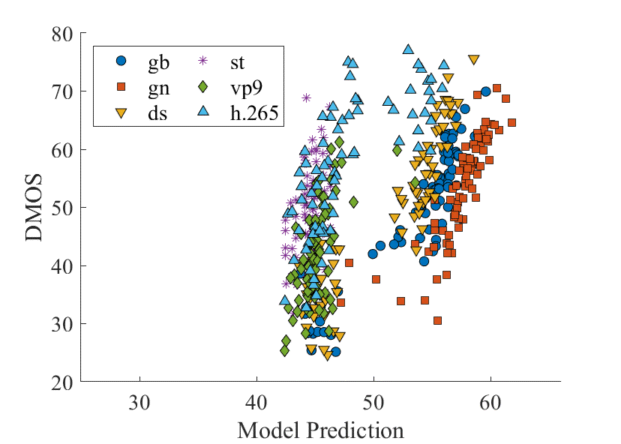}}
\caption{Scatter plots of all pairs of objective and subjective IQA scores using different IQA algorithms. 'gb' refers to Gaussian blur, 'gn' refers to Gaussian noise, 'ds' refers to downsampling, and 'st' refers to stitching.}
\label{fig:scatter_algo_dmos}
\end{figure*}

\section{Objective IQA Model Comparison}
\label{section5}
When evaluating the performance of IQA methods, we computed the IQA scores separately on the  left and right images, and used the average of these as the overall IQA score. Since BRISQUE requires training in advance, we split the database randomly, using 80\% of the data for training, and 20\% for testing. No contents were shared between training and testing. On each distortion type, BRISQUE was trained and tested using only features extracted on images having the corresponding distortion type, to allow measurement of the best case median performance, since performance of BRISQUE degrades in general when it has more distortions to measure. This process was done 1000 times and the median value was taken as the final IQA score. The IQA scores of the other methods were processed in the same way to avoid any bias. We tested and compared the following IQA models on our database. 
\begin{enumerate}
    \item \textit{Peak Signal-to-Noise Ratio (PSNR)} is the negative logarithm of the  pixel-wise mean squared error (MSE) function plus an additive offset between the reference and distorted images. 
    \item \textit{Weighted-to-Spherically-Uniform PSNR (WS-PSNR)}\cite{sun2017weighted} is a modification of PSNR that measures distortions in
    representation space and weights distortions according to the corresponding
    projection area in observation space. 
    \item \textit{Structural Similarity Index (SSIM)} is a widely used full reference image quality assessment model\cite{wang2004image} which captures local luminance, contrast, and structural information.
    \item \textit{Multiscale SSIM (MS-SSIM)} \cite{wang2003multiscale} is a variation of SSIM that captures quality information across multiple spatial scales.
    \item \textit{Visual Saliency-Induced Index (VSI)}\cite{zhang2014vsi} is a full reference visual saliency-based IQA method that also integrates gradient magnitude and chrominance features.
    \item \textit{Gradient Magnitude Similarity Deviation (GMSD)}\cite{xue2014gradient} is a simple gradient-based IQA method. It also uses spatial deviation pooling to aggregate the quality predictions.
    \item \textit{FSIM} \cite{zhang2011fsim} is a full reference IQA method that measures image quality based on local measurements of phase congruency and gradient magnitude.
    \item \textit{Mean Deviation Similarity Index (MDSI)}\cite{nafchi2016mean} is a full reference image quality evaluator that fuses gradient similarity, chromaticit, and deviation pooling features.
    \item \textit{Spherical Structural Similarity Index (S-SSIM)}\cite{chen2018spherical} is a weighted-to-spherically-uniform VR-IQA method which scales pixels with equal mapped spherical areas by equal factors when measuring distortion using SSIM. 
    \item \textit{BRISQUE}\cite{mittal2012no} is a NR IQA model that uses natural scene statistics features defined in the spatial domain.
    \item \textit{NIQE}\cite{mittal2013making} is a completely blind (unsupervised) image quality assessment model, in which the quality of a distorted image is computed in terms of its distance from a learned NSS model. 
\end{enumerate}
\subsection{Performance of Objective Methods} 
We tested the performance of the just-listed objective IQA models using three metrics: the Spearman's Rank Order Correlation Coefficient (SROCC), the Pearson Linear Correlation Coefficient (PLCC), and the Root Mean Square Error (RMSE). The SROCC assesses how well the relationship between an objective model prediction and human subjective scores can be described using a monotonic function. The PLCC measures the  accuracy of prediction of different objective models after performing a nonlinear logistic regression. We used a five-parameter logistic function:
\begin{equation}
    f(x) = \beta_1(\frac{1}{2} - \frac{1}{e^{\beta_2(x - \beta_3)}}) + \beta_4x + \beta_5
\label{nonlinear}
\end{equation}
where $x$ are the predicted scores, $f(x)$ is the mapped score, and $\beta_i (i = 1, 2, 3, 4, 5)$ are parameters to be fitted that minimize the mean squared error between the mapped scores and the subjective scores. The Root Mean Squared Error is the standard deviation of the  prediction errors. The performances of the compared IQA models is listed in Tables \ref{srocc_table}, \ref{plcc_table} and \ref{rmse_table}. In addition, scatter plots of all of the considered objective IQA models against DMOS are shown in Figure \ref{fig:scatter_algo_dmos}. 
\begin{table*}[]
\centering
\caption{SROCC of IQA Methods}
\label{srocc_table}
\begin{tabular}{|c|c|c|c|c|c|c|c|}
\hline
      &\scriptsize{OVERALL}&\scriptsize{GAUSSIAN BLUR}&\scriptsize{GAUSSIAN NOISE}&\scriptsize{DOWNSAMPLING}&\scriptsize{STITCHING}&\scriptsize{VP9}&\scriptsize{H.265}\\\hline
PSNR & 0.5791 & 0.7964 & 0.8964 & 0.8179 & 0.7929 & \textbf{0.5036} & 0.7750 \\\hline
WS-PSNR & 0.6371 & 0.8071 & \textbf{0.8893} & 0.8321 & 0.8000 & 0.6393 & 0.8571 \\\hline
SSIM & 0.6223 & \textbf{0.7750} & 0.9107 & 0.8286 & 0.5071 & 0.7607 & 0.7821 \\\hline
MS-SSIM & 0.7166 & 0.8536 & 0.9143 & 0.8000 & 0.7214 & 0.8107 & 0.9250 \\\hline
WS-SSIM & 0.6382 & 0.7964 & 0.9143 & 0.8571 & 0.5783 & 0.7500 & 0.8179 \\\hline
FSIM & 0.7007 & 0.9179 & 0.9143 & \textbf{0.7893} & \textbf{0.8179} & \textbf{0.8821} & 0.9357 \\\hline
VSI & 0.6776 & 0.9143 & 0.9143 & 0.7964 & 0.7964 & 0.8571 & 0.9429 \\\hline
GMSD & 0.7969 & 0.9000 & 0.9071 & 0.8214 & 0.7812 & 0.8429 & 0.9321 \\\hline
MDSI & 0.6945 & 0.9179 & \textbf{0.9179} & 0.7929 & 0.8143 & 0.8679 & 0.9429 \\\hline
BRISQUE & \textbf{0.8332} & \textbf{0.9464} & 0.8929 & \textbf{0.9464} & 0.5071 & 0.5893 & 0.7714 \\\hline
NIQE & \textbf{0.4608} & 0.9321 & 0.8929 & 0.8821 & \textbf{0.1107} & 0.5411 & \textbf{0.6786} \\\hline
\end{tabular}
\end{table*}

\begin{table*}[]
\centering
\caption{PLCC of IQA Methods}
\label{plcc_table}
\begin{tabular}{|c|c|c|c|c|c|c|c|}
\hline
      &\scriptsize{OVERALL}&\scriptsize{GAUSSIAN BLUR}&\scriptsize{GAUSSIAN NOISE}&\scriptsize{DOWNSAMPLING}&\scriptsize{STITCHING}&\scriptsize{VP9}&\scriptsize{H.265}\\\hline
PSNR & 0.6370 & 0.8707 & 0.9178 & 0.8448 & 0.7297 & 0.5034 & 0.7478 \\\hline
WS-PSNR & 0.6834 & 0.8726 & 0.9205 & 0.8535 & 0.7801 & 0.6526 & 0.8401 \\\hline
SSIM & 0.6792 & \textbf{0.8606} & 0.9360 & \textbf{0.8343} & 0.4810 & 0.7227 & 0.7433 \\\hline
MS-SSIM & 0.7438 & 0.9076 & 0.9329 & 0.8865 & 0.6945 & 0.8288 & 0.9169 \\\hline
WS-SSIM & 0.6961 & 0.8761 & \textbf{0.9416} & 0.8630 & 0.5675 & 0.7358 & 0.7876 \\\hline
FSIM & 0.7555 & 0.9488 & 0.9379 & 0.9096 & \textbf{0.8046} & \textbf{0.8875} & 0.9375 \\\hline
VSI & 0.7274 & 0.9465 & 0.9311 & 0.9021 & 0.7216 & 0.8634 & \textbf{0.9404} \\\hline
GMSD & 0.8067 & 0.9409 & 0.9289 & 0.9088 & 0.7653 & 0.8622 & 0.9335 \\\hline
MDSI & 0.7412 & 0.9505 & 0.9369 & 0.9221 & 0.7983 & 0.8777 & 0.9399 \\\hline
BRISQUE & \textbf{0.8280} & \textbf{0.9543} & 0.9359 & \textbf{0.9533} & 0.5093 & 0.6033 & 0.7401 \\\hline
NIQE & \textbf{0.4646} & 0.9479 & \textbf{0.9099} & 0.9192 & \textbf{0.0233} & \textbf{0.4774} & \textbf{0.6499} \\\hline

\end{tabular}
\end{table*}

\begin{table*}[]
\centering
\caption{RMSE of IQA Methods}
\label{rmse_table}
\begin{tabular}{|c|c|c|c|c|c|c|c|}
\hline
      &\scriptsize{OVERALL}&\scriptsize{GAUSSIAN BLUR}&\scriptsize{GAUSSIAN NOISE}&\scriptsize{DOWNSAMPLING}&\scriptsize{STITCHING}&\scriptsize{VP9}&\scriptsize{H.265}\\\hline   
PSNR & 9.0900 & \textbf{7.079} & 4.1236 & 7.7767 & 5.9731 & 8.0565 & 8.9386 \\\hline
WS-PSNR & 8.6951 & 6.7828 & 4.2306 & 7.6926 & 5.4808 & 7.5422 & 7.4981 \\\hline
SSIM & 8.8417 & 6.4503 & 3.7293 & \textbf{7.9927} & 7.8591 & 6.9617 & 9.1335 \\\hline
MS-SSIM & 7.8740 & 5.5060 & 3.9153 & 6.6031 & 6.4280 & 5.4026 & 5.5419 \\\hline
S-SSIM & 8.5785 & 6.1632 & 3.6075 & 7.4096 & 7.5257 & 6.8337 & 8.2770 \\\hline
FSIM & 7.5572 & 4.1814 & 3.5923 & 5.7645 & \textbf{5.2133} & \textbf{4.6247} & 5.0247 \\\hline
VSI & 7.9557 & 4.3320 & 3.7436 & 6.0238 & 6.0504 & 5.1298 & 4.7766 \\\hline
GMSD & 6.9385 & 4.5461 & 3.8439 & 6.1944 & 5.6549 & 5.1752 & 4.9873 \\\hline
MDSI & 7.7672 & 4.3169 & \textbf{3.5571} & 5.6266 & 5.3215 & 4.7613 & \textbf{4.7147} \\\hline
BRISQUE & \textbf{6.6161} & \textbf{3.8550} & 3.6010 & \textbf{4.8176} & 6.9558 & 7.8166 & 9.3263 \\\hline
NIQE & \textbf{10.0793} & 4.3405 & \textbf{4.5707} & 6.0460 & \textbf{7.9863} & \textbf{8.1521} & \textbf{10.3686} \\\hline   

\end{tabular}
\end{table*}

\subsection{Statistical Evaluation}
To evaluate whether two IQA methods are significantly different, we performed an F-test on the residuals between the IQA scores after non-linear mapping and the DMOS\cite{sheikh2006statistical}. The assumption is that the two sets of residuals are Gaussian with zero means. Thus, to test whether they come from the same distribution depends on whether they have the same variance. The null hypothesis is that the residuals from one IQA come from the same distribution and are statistically indistinguishable from the residuals from another IQA. Each entry in the table consists of 6 symbols. A value of `1' in the table represents that the row algorithm is statistically superior to the column algorithm, while a value of `0' means the opposite.
A value of `-' indicates that the row and column algorithms are statistically indistinguishable (or equivalent). The position of the symbols corresponds to the following datasets: Gaussian blur, Gaussian noise, downsampling, stitching, VP9, H.265, and all data. The results are shown in 
Table \ref{f-test}.

\begin{sidewaystable}[]
\centering
\caption{Statistical Significance Matrix based on IQA-DMOS residuals. All statistical tests are performed at 95\% confidence.}
\label{f-test}
\begin{tabular}{|c|c|c|c|c|c|c|c|c|c|c|c|}\hline
& PSNR & WS-PSNR & SSIM & MS-SSIM & S-SSIM & FSIM & VSI & GMSD & MDSI & BRISQUE & NIQE \\ \hline
PSNR & -\hspace{1pt}-\hspace{1pt}-\hspace{1pt}-\hspace{1pt}-\hspace{1pt}-\hspace{1pt}-\hspace{1pt} & 0\hspace{1pt}-\hspace{1pt}-\hspace{1pt}0\hspace{1pt}-\hspace{1pt}0\hspace{1pt}0\hspace{1pt} & 0\hspace{1pt}0\hspace{1pt}1\hspace{1pt}1\hspace{1pt}0\hspace{1pt}0\hspace{1pt}0\hspace{1pt} & 0\hspace{1pt}0\hspace{1pt}0\hspace{1pt}1\hspace{1pt}0\hspace{1pt}0\hspace{1pt}0\hspace{1pt} & 0\hspace{1pt}0\hspace{1pt}0\hspace{1pt}1\hspace{1pt}0\hspace{1pt}0\hspace{1pt}0\hspace{1pt} & 0\hspace{1pt}0\hspace{1pt}0\hspace{1pt}0\hspace{1pt}0\hspace{1pt}0\hspace{1pt}0\hspace{1pt} & 0\hspace{1pt}0\hspace{1pt}0\hspace{1pt}0\hspace{1pt}0\hspace{1pt}0\hspace{1pt}0\hspace{1pt} & 0\hspace{1pt}0\hspace{1pt}0\hspace{1pt}0\hspace{1pt}0\hspace{1pt}0\hspace{1pt}0\hspace{1pt} & 0\hspace{1pt}0\hspace{1pt}0\hspace{1pt}0\hspace{1pt}0\hspace{1pt}0\hspace{1pt}0\hspace{1pt} & 0\hspace{1pt}0\hspace{1pt}0\hspace{1pt}1\hspace{1pt}1\hspace{1pt}-\hspace{1pt}0\hspace{1pt} & 0\hspace{1pt}1\hspace{1pt}0\hspace{1pt}1\hspace{1pt}0\hspace{1pt}1\hspace{1pt}1\hspace{1pt} \\\hline
WS-PSNR & 1\hspace{1pt}-\hspace{1pt}-\hspace{1pt}1\hspace{1pt}-\hspace{1pt}1\hspace{1pt}1\hspace{1pt} & -\hspace{1pt}-\hspace{1pt}-\hspace{1pt}-\hspace{1pt}-\hspace{1pt}-\hspace{1pt}-\hspace{1pt} & 0\hspace{1pt}0\hspace{1pt}1\hspace{1pt}1\hspace{1pt}0\hspace{1pt}1\hspace{1pt}1\hspace{1pt} & 0\hspace{1pt}0\hspace{1pt}0\hspace{1pt}1\hspace{1pt}0\hspace{1pt}0\hspace{1pt}0\hspace{1pt} & 0\hspace{1pt}0\hspace{1pt}-\hspace{1pt}1\hspace{1pt}0\hspace{1pt}1\hspace{1pt}0\hspace{1pt} & 0\hspace{1pt}0\hspace{1pt}0\hspace{1pt}0\hspace{1pt}0\hspace{1pt}0\hspace{1pt}0\hspace{1pt} & 0\hspace{1pt}0\hspace{1pt}0\hspace{1pt}0\hspace{1pt}0\hspace{1pt}0\hspace{1pt}0\hspace{1pt} & 0\hspace{1pt}0\hspace{1pt}0\hspace{1pt}0\hspace{1pt}0\hspace{1pt}0\hspace{1pt}0\hspace{1pt} & 0\hspace{1pt}0\hspace{1pt}0\hspace{1pt}0\hspace{1pt}0\hspace{1pt}0\hspace{1pt}0\hspace{1pt} & 0\hspace{1pt}0\hspace{1pt}0\hspace{1pt}1\hspace{1pt}1\hspace{1pt}1\hspace{1pt}0\hspace{1pt} & 0\hspace{1pt}1\hspace{1pt}0\hspace{1pt}1\hspace{1pt}0\hspace{1pt}1\hspace{1pt}1\hspace{1pt} \\\hline
SSIM & 1\hspace{1pt}1\hspace{1pt}0\hspace{1pt}0\hspace{1pt}1\hspace{1pt}1\hspace{1pt}1\hspace{1pt} & 1\hspace{1pt}1\hspace{1pt}0\hspace{1pt}0\hspace{1pt}1\hspace{1pt}0\hspace{1pt}0\hspace{1pt} & -\hspace{1pt}-\hspace{1pt}-\hspace{1pt}-\hspace{1pt}-\hspace{1pt}-\hspace{1pt}-\hspace{1pt} & 0\hspace{1pt}0\hspace{1pt}0\hspace{1pt}0\hspace{1pt}0\hspace{1pt}0\hspace{1pt}0\hspace{1pt} & 0\hspace{1pt}0\hspace{1pt}0\hspace{1pt}1\hspace{1pt}0\hspace{1pt}0\hspace{1pt}0\hspace{1pt} & 0\hspace{1pt}0\hspace{1pt}0\hspace{1pt}0\hspace{1pt}0\hspace{1pt}0\hspace{1pt}0\hspace{1pt} & 0\hspace{1pt}0\hspace{1pt}0\hspace{1pt}0\hspace{1pt}0\hspace{1pt}0\hspace{1pt}0\hspace{1pt} & 0\hspace{1pt}-\hspace{1pt}0\hspace{1pt}0\hspace{1pt}0\hspace{1pt}0\hspace{1pt}0\hspace{1pt} & 0\hspace{1pt}0\hspace{1pt}0\hspace{1pt}0\hspace{1pt}0\hspace{1pt}0\hspace{1pt}0\hspace{1pt} & 0\hspace{1pt}0\hspace{1pt}0\hspace{1pt}0\hspace{1pt}1\hspace{1pt}1\hspace{1pt}0\hspace{1pt} & 0\hspace{1pt}1\hspace{1pt}0\hspace{1pt}0\hspace{1pt}1\hspace{1pt}1\hspace{1pt}1\hspace{1pt} \\\hline
MSSSIM & 1\hspace{1pt}1\hspace{1pt}1\hspace{1pt}0\hspace{1pt}1\hspace{1pt}1\hspace{1pt}1\hspace{1pt} & 1\hspace{1pt}1\hspace{1pt}1\hspace{1pt}0\hspace{1pt}1\hspace{1pt}1\hspace{1pt}1\hspace{1pt} & 1\hspace{1pt}1\hspace{1pt}1\hspace{1pt}1\hspace{1pt}1\hspace{1pt}1\hspace{1pt}1\hspace{1pt} & -\hspace{1pt}-\hspace{1pt}-\hspace{1pt}-\hspace{1pt}-\hspace{1pt}-\hspace{1pt}-\hspace{1pt} & 1\hspace{1pt}0\hspace{1pt}1\hspace{1pt}1\hspace{1pt}1\hspace{1pt}1\hspace{1pt}1\hspace{1pt} & 0\hspace{1pt}0\hspace{1pt}0\hspace{1pt}0\hspace{1pt}0\hspace{1pt}0\hspace{1pt}0\hspace{1pt} & 0\hspace{1pt}0\hspace{1pt}0\hspace{1pt}0\hspace{1pt}0\hspace{1pt}0\hspace{1pt}1\hspace{1pt} & 0\hspace{1pt}1\hspace{1pt}0\hspace{1pt}0\hspace{1pt}0\hspace{1pt}0\hspace{1pt}0\hspace{1pt} & 0\hspace{1pt}0\hspace{1pt}0\hspace{1pt}0\hspace{1pt}0\hspace{1pt}0\hspace{1pt}0\hspace{1pt} & 0\hspace{1pt}0\hspace{1pt}0\hspace{1pt}1\hspace{1pt}1\hspace{1pt}1\hspace{1pt}0\hspace{1pt} & 0\hspace{1pt}1\hspace{1pt}0\hspace{1pt}1\hspace{1pt}1\hspace{1pt}1\hspace{1pt}1\hspace{1pt} \\\hline
S-SSIM & 1\hspace{1pt}1\hspace{1pt}1\hspace{1pt}0\hspace{1pt}1\hspace{1pt}1\hspace{1pt}1\hspace{1pt} & 1\hspace{1pt}1\hspace{1pt}-\hspace{1pt}0\hspace{1pt}1\hspace{1pt}0\hspace{1pt}1\hspace{1pt} & 1\hspace{1pt}1\hspace{1pt}1\hspace{1pt}0\hspace{1pt}1\hspace{1pt}1\hspace{1pt}1\hspace{1pt} & 0\hspace{1pt}1\hspace{1pt}0\hspace{1pt}0\hspace{1pt}0\hspace{1pt}0\hspace{1pt}0\hspace{1pt} & -\hspace{1pt}-\hspace{1pt}-\hspace{1pt}-\hspace{1pt}-\hspace{1pt}-\hspace{1pt}-\hspace{1pt} & 0\hspace{1pt}0\hspace{1pt}0\hspace{1pt}0\hspace{1pt}0\hspace{1pt}0\hspace{1pt}0\hspace{1pt} & 0\hspace{1pt}-\hspace{1pt}0\hspace{1pt}0\hspace{1pt}0\hspace{1pt}0\hspace{1pt}0\hspace{1pt} & 0\hspace{1pt}1\hspace{1pt}0\hspace{1pt}0\hspace{1pt}0\hspace{1pt}0\hspace{1pt}0\hspace{1pt} & 0\hspace{1pt}0\hspace{1pt}0\hspace{1pt}0\hspace{1pt}0\hspace{1pt}0\hspace{1pt}0\hspace{1pt} & 0\hspace{1pt}0\hspace{1pt}0\hspace{1pt}0\hspace{1pt}1\hspace{1pt}1\hspace{1pt}0\hspace{1pt} & 0\hspace{1pt}1\hspace{1pt}0\hspace{1pt}0\hspace{1pt}1\hspace{1pt}1\hspace{1pt}1\hspace{1pt} \\\hline
FSIM & 1\hspace{1pt}1\hspace{1pt}1\hspace{1pt}1\hspace{1pt}1\hspace{1pt}1\hspace{1pt}1\hspace{1pt} & 1\hspace{1pt}1\hspace{1pt}1\hspace{1pt}1\hspace{1pt}1\hspace{1pt}1\hspace{1pt}1\hspace{1pt} & 1\hspace{1pt}1\hspace{1pt}1\hspace{1pt}1\hspace{1pt}1\hspace{1pt}1\hspace{1pt}1\hspace{1pt} & 1\hspace{1pt}1\hspace{1pt}1\hspace{1pt}1\hspace{1pt}1\hspace{1pt}1\hspace{1pt}1\hspace{1pt} & 1\hspace{1pt}1\hspace{1pt}1\hspace{1pt}1\hspace{1pt}1\hspace{1pt}1\hspace{1pt}1\hspace{1pt} & -\hspace{1pt}-\hspace{1pt}-\hspace{1pt}-\hspace{1pt}-\hspace{1pt}-\hspace{1pt}-\hspace{1pt} & -\hspace{1pt}1\hspace{1pt}-\hspace{1pt}1\hspace{1pt}1\hspace{1pt}0\hspace{1pt}1\hspace{1pt} & 1\hspace{1pt}1\hspace{1pt}-\hspace{1pt}1\hspace{1pt}1\hspace{1pt}0\hspace{1pt}0\hspace{1pt} & 1\hspace{1pt}-\hspace{1pt}0\hspace{1pt}1\hspace{1pt}-\hspace{1pt}0\hspace{1pt}1\hspace{1pt} & 0\hspace{1pt}-\hspace{1pt}0\hspace{1pt}1\hspace{1pt}1\hspace{1pt}1\hspace{1pt}0\hspace{1pt} & 0\hspace{1pt}1\hspace{1pt}0\hspace{1pt}1\hspace{1pt}1\hspace{1pt}1\hspace{1pt}1\hspace{1pt} \\\hline
VSI & 1\hspace{1pt}1\hspace{1pt}1\hspace{1pt}1\hspace{1pt}1\hspace{1pt}1\hspace{1pt}1\hspace{1pt} & 1\hspace{1pt}1\hspace{1pt}1\hspace{1pt}1\hspace{1pt}1\hspace{1pt}1\hspace{1pt}1\hspace{1pt} & 1\hspace{1pt}1\hspace{1pt}1\hspace{1pt}1\hspace{1pt}1\hspace{1pt}1\hspace{1pt}1\hspace{1pt} & 1\hspace{1pt}1\hspace{1pt}1\hspace{1pt}1\hspace{1pt}1\hspace{1pt}1\hspace{1pt}0\hspace{1pt} & 1\hspace{1pt}-\hspace{1pt}1\hspace{1pt}1\hspace{1pt}1\hspace{1pt}1\hspace{1pt}1\hspace{1pt} & -\hspace{1pt}0\hspace{1pt}-\hspace{1pt}0\hspace{1pt}0\hspace{1pt}1\hspace{1pt}0\hspace{1pt} & -\hspace{1pt}-\hspace{1pt}-\hspace{1pt}-\hspace{1pt}-\hspace{1pt}-\hspace{1pt}-\hspace{1pt} & 1\hspace{1pt}1\hspace{1pt}-\hspace{1pt}-\hspace{1pt}-\hspace{1pt}-\hspace{1pt}0\hspace{1pt} & -\hspace{1pt}0\hspace{1pt}0\hspace{1pt}0\hspace{1pt}0\hspace{1pt}0\hspace{1pt}0\hspace{1pt} & 0\hspace{1pt}0\hspace{1pt}0\hspace{1pt}1\hspace{1pt}1\hspace{1pt}1\hspace{1pt}0\hspace{1pt} & 0\hspace{1pt}1\hspace{1pt}0\hspace{1pt}1\hspace{1pt}1\hspace{1pt}1\hspace{1pt}1\hspace{1pt} \\\hline
GMSD & 1\hspace{1pt}1\hspace{1pt}1\hspace{1pt}1\hspace{1pt}1\hspace{1pt}1\hspace{1pt}1\hspace{1pt} & 1\hspace{1pt}1\hspace{1pt}1\hspace{1pt}1\hspace{1pt}1\hspace{1pt}1\hspace{1pt}1\hspace{1pt} & 1\hspace{1pt}-\hspace{1pt}1\hspace{1pt}1\hspace{1pt}1\hspace{1pt}1\hspace{1pt}1\hspace{1pt} & 1\hspace{1pt}0\hspace{1pt}1\hspace{1pt}1\hspace{1pt}1\hspace{1pt}1\hspace{1pt}1\hspace{1pt} & 1\hspace{1pt}0\hspace{1pt}1\hspace{1pt}1\hspace{1pt}1\hspace{1pt}1\hspace{1pt}1\hspace{1pt} & 0\hspace{1pt}0\hspace{1pt}-\hspace{1pt}0\hspace{1pt}0\hspace{1pt}1\hspace{1pt}1\hspace{1pt} & 0\hspace{1pt}0\hspace{1pt}-\hspace{1pt}-\hspace{1pt}-\hspace{1pt}-\hspace{1pt}1\hspace{1pt} & -\hspace{1pt}-\hspace{1pt}-\hspace{1pt}-\hspace{1pt}-\hspace{1pt}-\hspace{1pt}-\hspace{1pt} & -\hspace{1pt}0\hspace{1pt}0\hspace{1pt}0\hspace{1pt}0\hspace{1pt}0\hspace{1pt}1\hspace{1pt} & 0\hspace{1pt}0\hspace{1pt}0\hspace{1pt}1\hspace{1pt}1\hspace{1pt}1\hspace{1pt}0\hspace{1pt} & 0\hspace{1pt}1\hspace{1pt}0\hspace{1pt}1\hspace{1pt}1\hspace{1pt}1\hspace{1pt}1\hspace{1pt} \\\hline
MDSI & 1\hspace{1pt}1\hspace{1pt}1\hspace{1pt}1\hspace{1pt}1\hspace{1pt}1\hspace{1pt}1\hspace{1pt} & 1\hspace{1pt}1\hspace{1pt}1\hspace{1pt}1\hspace{1pt}1\hspace{1pt}1\hspace{1pt}1\hspace{1pt} & 1\hspace{1pt}1\hspace{1pt}1\hspace{1pt}1\hspace{1pt}1\hspace{1pt}1\hspace{1pt}1\hspace{1pt} & 1\hspace{1pt}1\hspace{1pt}1\hspace{1pt}1\hspace{1pt}1\hspace{1pt}1\hspace{1pt}1\hspace{1pt} & 1\hspace{1pt}1\hspace{1pt}1\hspace{1pt}1\hspace{1pt}1\hspace{1pt}1\hspace{1pt}1\hspace{1pt} & 0\hspace{1pt}-\hspace{1pt}1\hspace{1pt}0\hspace{1pt}-\hspace{1pt}1\hspace{1pt}0\hspace{1pt} & -\hspace{1pt}1\hspace{1pt}1\hspace{1pt}1\hspace{1pt}1\hspace{1pt}1\hspace{1pt}1\hspace{1pt} & -\hspace{1pt}1\hspace{1pt}1\hspace{1pt}1\hspace{1pt}1\hspace{1pt}1\hspace{1pt}0\hspace{1pt} & -\hspace{1pt}-\hspace{1pt}-\hspace{1pt}-\hspace{1pt}-\hspace{1pt}-\hspace{1pt}-\hspace{1pt} & 0\hspace{1pt}-\hspace{1pt}0\hspace{1pt}1\hspace{1pt}1\hspace{1pt}1\hspace{1pt}0\hspace{1pt} & 0\hspace{1pt}1\hspace{1pt}1\hspace{1pt}1\hspace{1pt}1\hspace{1pt}1\hspace{1pt}1\hspace{1pt} \\\hline
BRISQUE & 1\hspace{1pt}1\hspace{1pt}1\hspace{1pt}0\hspace{1pt}0\hspace{1pt}-\hspace{1pt}1\hspace{1pt} & 1\hspace{1pt}1\hspace{1pt}1\hspace{1pt}0\hspace{1pt}0\hspace{1pt}0\hspace{1pt}1\hspace{1pt} & 1\hspace{1pt}1\hspace{1pt}1\hspace{1pt}1\hspace{1pt}0\hspace{1pt}0\hspace{1pt}1\hspace{1pt} & 1\hspace{1pt}1\hspace{1pt}1\hspace{1pt}0\hspace{1pt}0\hspace{1pt}0\hspace{1pt}1\hspace{1pt} & 1\hspace{1pt}1\hspace{1pt}1\hspace{1pt}1\hspace{1pt}0\hspace{1pt}0\hspace{1pt}1\hspace{1pt} & 1\hspace{1pt}-\hspace{1pt}1\hspace{1pt}0\hspace{1pt}0\hspace{1pt}0\hspace{1pt}1\hspace{1pt} & 1\hspace{1pt}1\hspace{1pt}1\hspace{1pt}0\hspace{1pt}0\hspace{1pt}0\hspace{1pt}1\hspace{1pt} & 1\hspace{1pt}1\hspace{1pt}1\hspace{1pt}0\hspace{1pt}0\hspace{1pt}0\hspace{1pt}1\hspace{1pt} & 1\hspace{1pt}-\hspace{1pt}1\hspace{1pt}0\hspace{1pt}0\hspace{1pt}0\hspace{1pt}1\hspace{1pt} & -\hspace{1pt}-\hspace{1pt}-\hspace{1pt}-\hspace{1pt}-\hspace{1pt}-\hspace{1pt}-\hspace{1pt} & 1\hspace{1pt}1\hspace{1pt}1\hspace{1pt}1\hspace{1pt}0\hspace{1pt}1\hspace{1pt}1\hspace{1pt} \\\hline
NIQE & 1\hspace{1pt}0\hspace{1pt}1\hspace{1pt}0\hspace{1pt}1\hspace{1pt}0\hspace{1pt}0\hspace{1pt} & 1\hspace{1pt}0\hspace{1pt}1\hspace{1pt}0\hspace{1pt}1\hspace{1pt}0\hspace{1pt}0\hspace{1pt} & 1\hspace{1pt}0\hspace{1pt}1\hspace{1pt}1\hspace{1pt}0\hspace{1pt}0\hspace{1pt}0\hspace{1pt} & 1\hspace{1pt}0\hspace{1pt}1\hspace{1pt}0\hspace{1pt}0\hspace{1pt}0\hspace{1pt}0\hspace{1pt} & 1\hspace{1pt}0\hspace{1pt}1\hspace{1pt}1\hspace{1pt}0\hspace{1pt}0\hspace{1pt}0\hspace{1pt} & 1\hspace{1pt}0\hspace{1pt}1\hspace{1pt}0\hspace{1pt}0\hspace{1pt}0\hspace{1pt}0\hspace{1pt} & 1\hspace{1pt}0\hspace{1pt}1\hspace{1pt}0\hspace{1pt}0\hspace{1pt}0\hspace{1pt}0\hspace{1pt} & 1\hspace{1pt}0\hspace{1pt}1\hspace{1pt}0\hspace{1pt}0\hspace{1pt}0\hspace{1pt}0\hspace{1pt} & 1\hspace{1pt}0\hspace{1pt}0\hspace{1pt}0\hspace{1pt}0\hspace{1pt}0\hspace{1pt}0\hspace{1pt} & 0\hspace{1pt}0\hspace{1pt}0\hspace{1pt}0\hspace{1pt}1\hspace{1pt}0\hspace{1pt}0\hspace{1pt} & -\hspace{1pt}-\hspace{1pt}-\hspace{1pt}-\hspace{1pt}-\hspace{1pt}-\hspace{1pt}-\hspace{1pt} \\ \hline
\end{tabular} 
\end{sidewaystable}

\subsection{Analysis of Eye Tracking Data}
\begin{figure*} [!ht]
\centerline{
\subfigure[]{
   \label{DMOS}
\includegraphics[width=0.4\columnwidth]{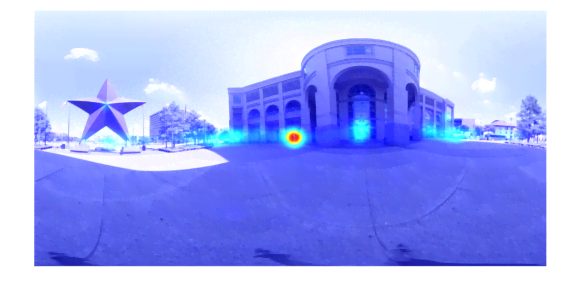}}
\subfigure[]{
   \label{SROCC}
\includegraphics[width=0.4\columnwidth]{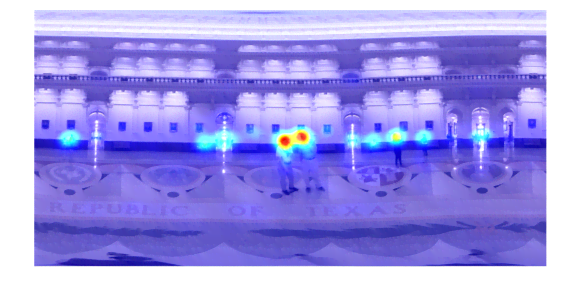}}
\subfigure[]{
   \label{SROCC}
\includegraphics[width=0.4\columnwidth]{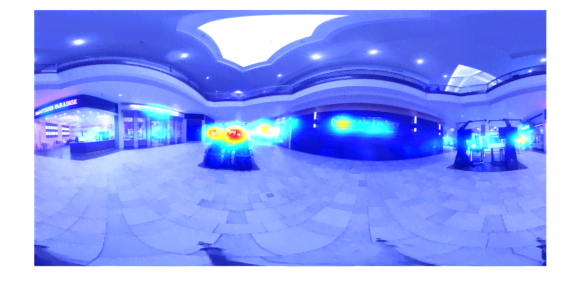}}
\subfigure[]{
   \label{SROCC}
\includegraphics[width=0.4\columnwidth]{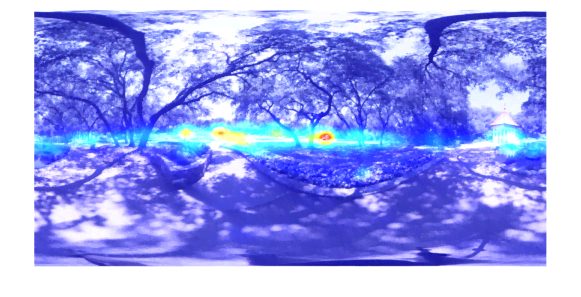}}
\subfigure[]{
   \label{gaze_e}
\includegraphics[width=0.4\columnwidth]{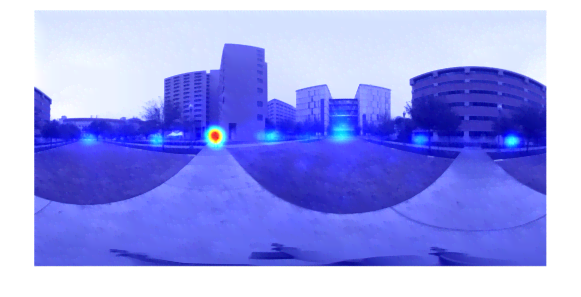}}
}
\centerline{
\subfigure[]{
   \label{DMOS}
\includegraphics[width=0.4\columnwidth]{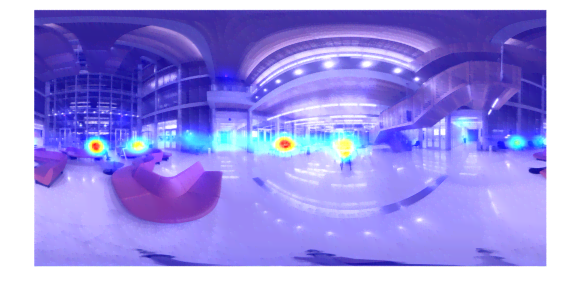}} 
\subfigure[]{
   \label{SROCC}
\includegraphics[width=0.4\columnwidth]{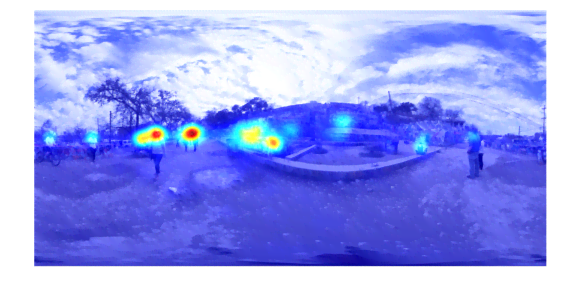}}
\subfigure[]{
   \label{SROCC}
\includegraphics[width=0.4\columnwidth]{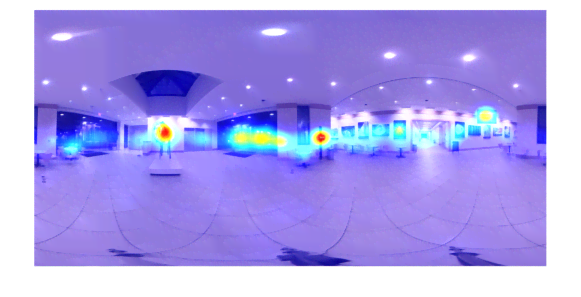}}
\subfigure[]{
   \label{SROCC}
\includegraphics[width=0.4\columnwidth]{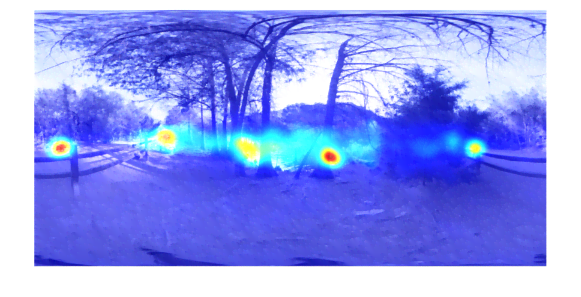}}
\subfigure[]{
   \label{gaze_j}
\includegraphics[width=0.4\columnwidth]{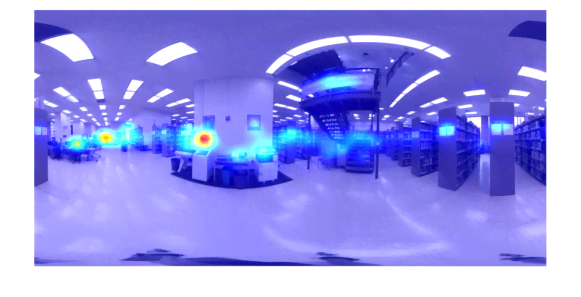}}
}
\centerline{
\subfigure[]{
   \label{DMOS}
\includegraphics[width=0.4\columnwidth]{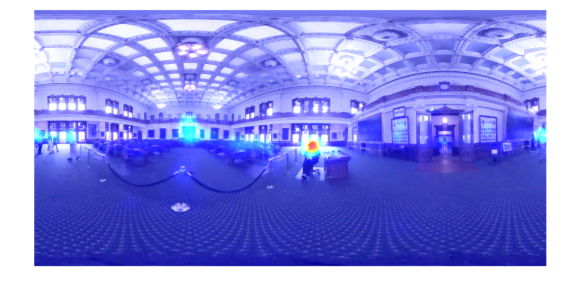}} 
\subfigure[]{
   \label{SROCC}
\includegraphics[width=0.4\columnwidth]{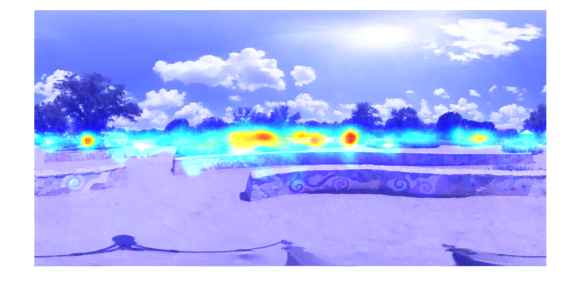}}
\subfigure[]{
   \label{SROCC}
\includegraphics[width=0.4\columnwidth]{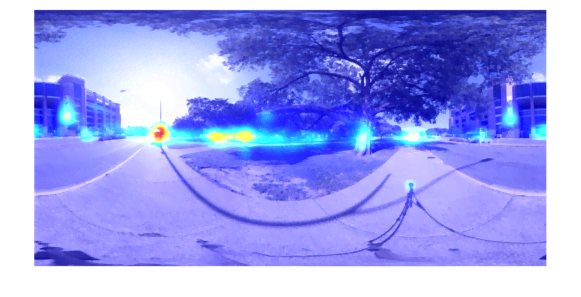}}
\subfigure[]{
   \label{SROCC}
\includegraphics[width=0.4\columnwidth]{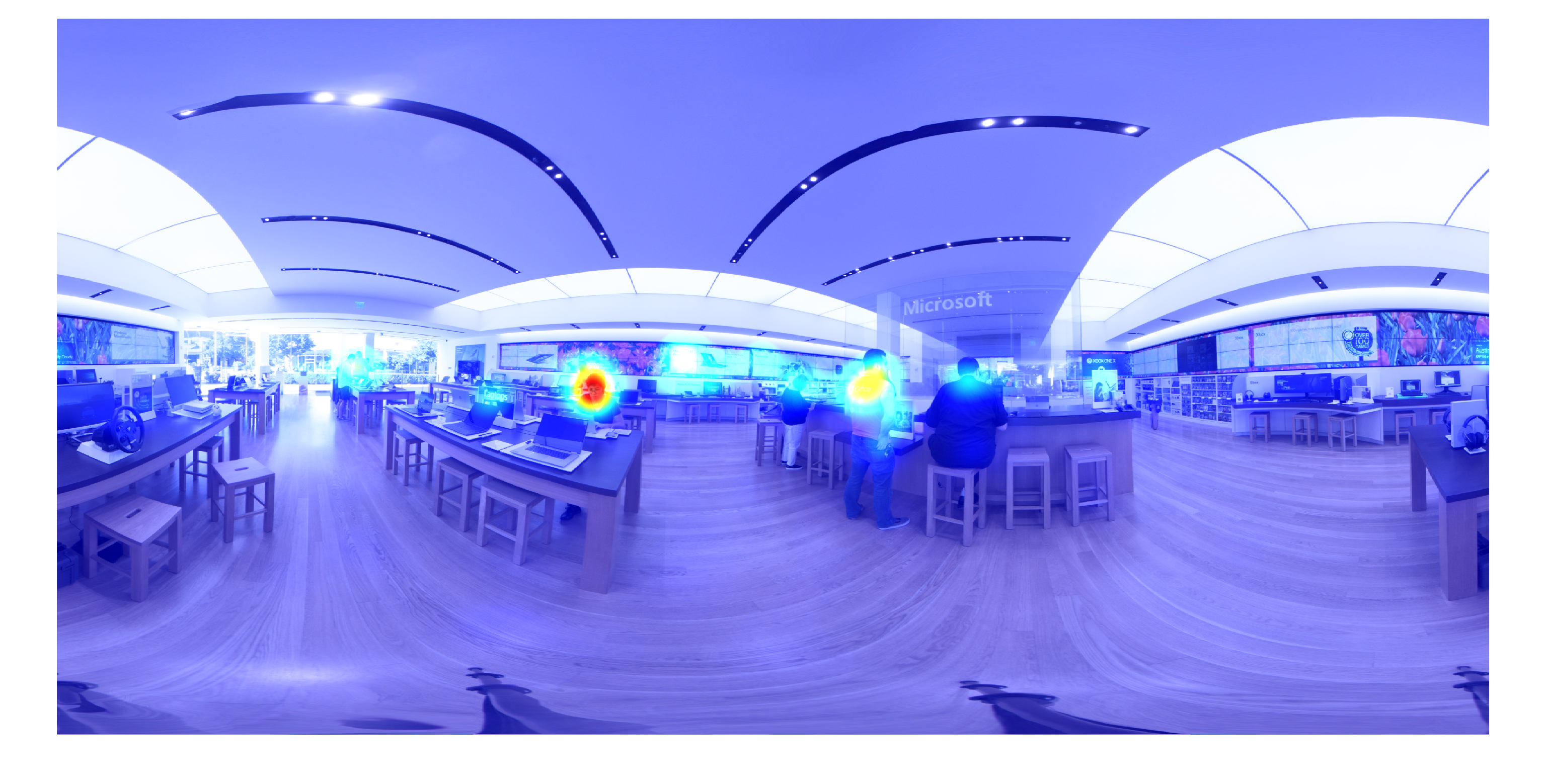}}
\subfigure[]{
   \label{SROCC}
\includegraphics[width=0.4\columnwidth]{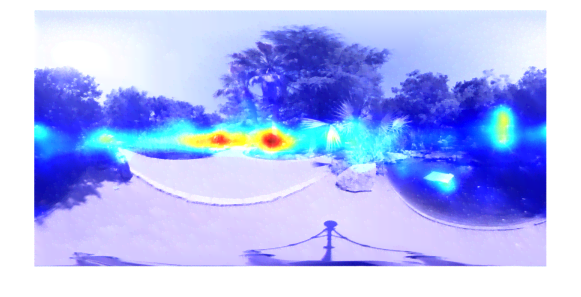}}
}
\caption{Example gaze maps}
\label{fig:db_gazemaps}
\end{figure*}

\begin{figure*} [!ht]
\centerline{
\subfigure[Longitude (x axis)]{
   \label{longitude}
\includegraphics[width=0.8\columnwidth]{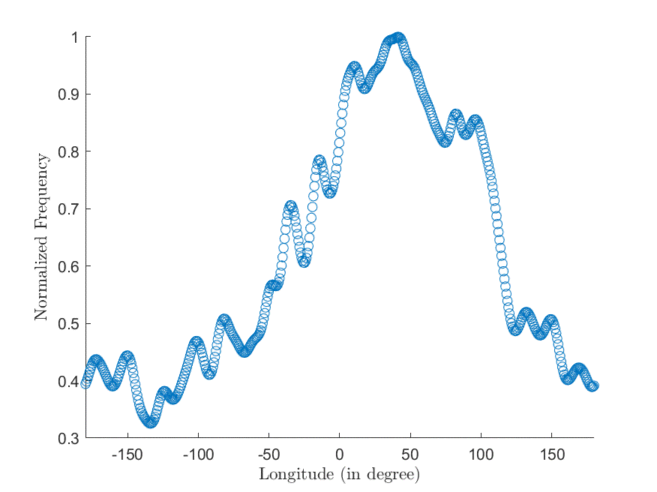}}
\subfigure[Latitude (y axis)]{
   \label{latitude}
\includegraphics[width=0.8\columnwidth]{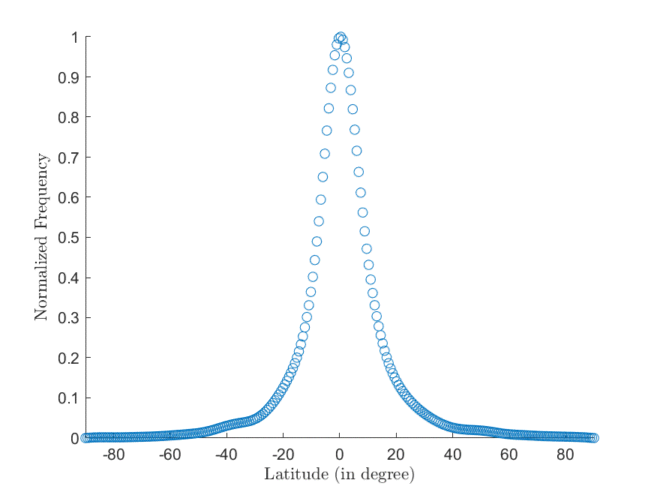}} 
}
\caption{Frequency of viewing directions.}
\label{fig:latitude_longitude}
\end{figure*}

\begin{figure*} [!ht]
\centerline{
\subfigure[]{
   \label{DMOS}
\includegraphics[width=0.5\columnwidth]{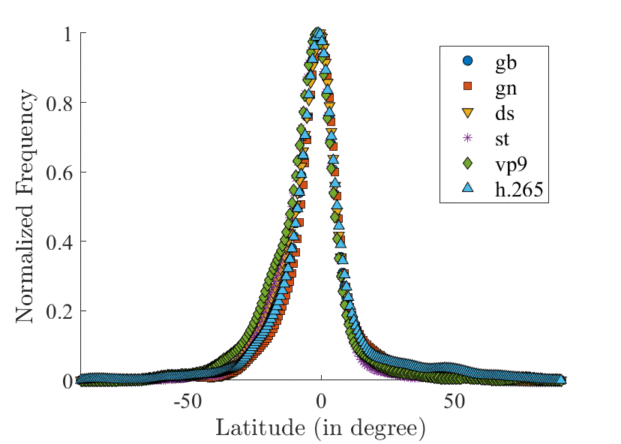}}
\subfigure[]{
   \label{SROCC}
\includegraphics[width=0.5\columnwidth]{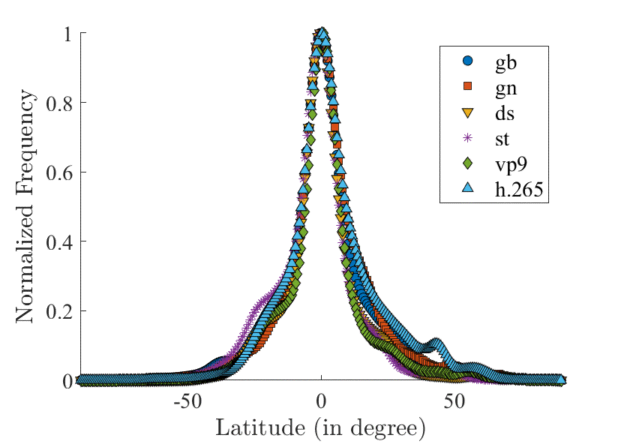}}
\subfigure[]{
   \label{SROCC}
\includegraphics[width=0.5\columnwidth]{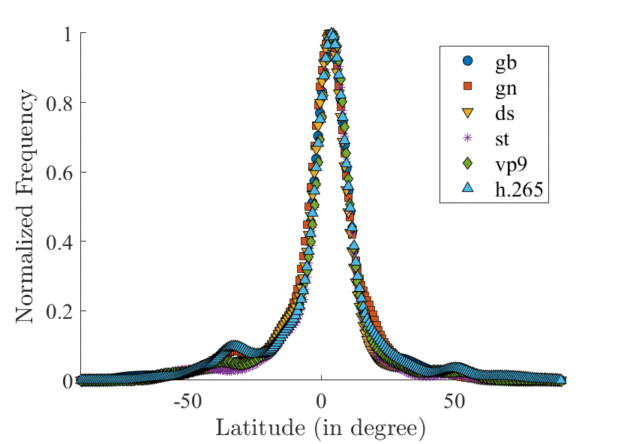}}
\subfigure[]{
   \label{SROCC}
\includegraphics[width=0.5\columnwidth]{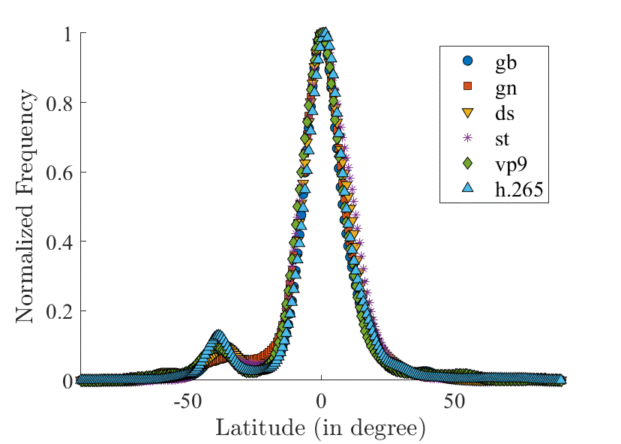}}
}
\caption{Example frequency plots of latitude viewing directions for four exemplar contents.}
\label{fig:latcontent_gazemaps}
\end{figure*}

\begin{figure*} [!ht]
\centerline{
\subfigure[]{
   \label{longitude_cheese}
\includegraphics[width=0.5\columnwidth]{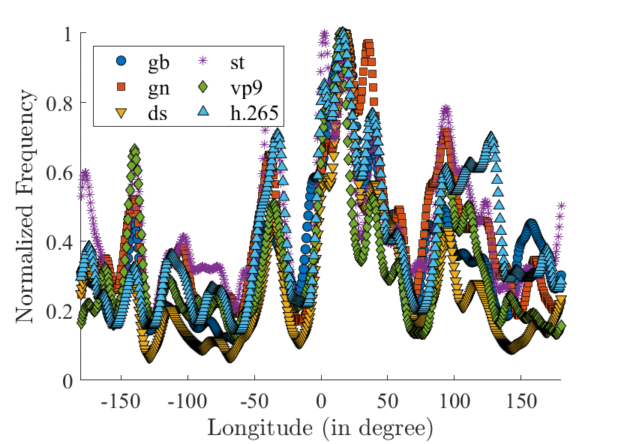}}
\subfigure[]{
   \label{SROCC}
\includegraphics[width=0.5\columnwidth]{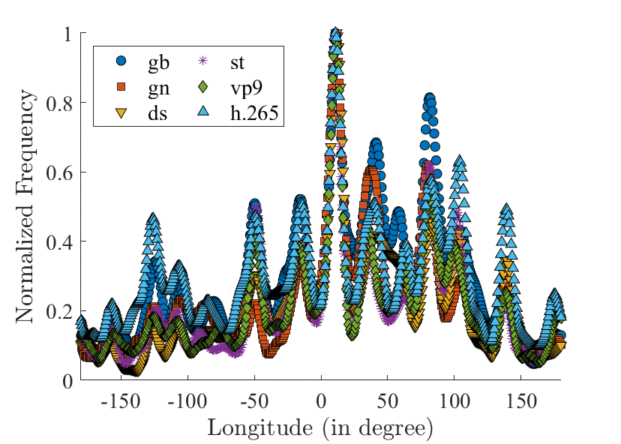}}
\subfigure[]{
   \label{SROCC}
\includegraphics[width=0.5\columnwidth]{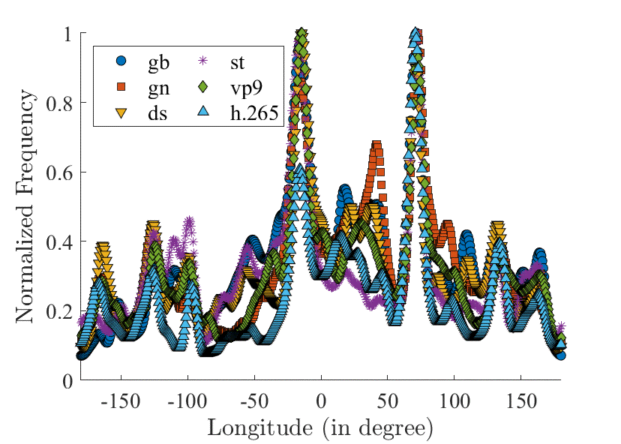}}
\subfigure[]{
   \label{SROCC}
\includegraphics[width=0.5\columnwidth]{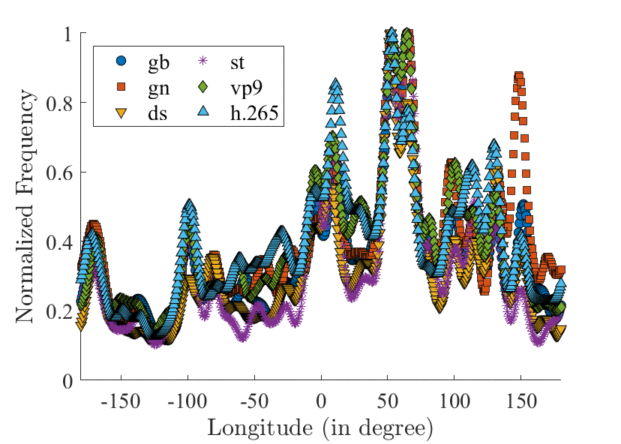}}
}
\caption{Example frequency plots of longitude viewing directions for four contents.}
\label{fig:longcontent_gazemaps}
\end{figure*}

\begin{figure*} [!ht]
\centerline{
\subfigure[Gaussian Blur]{
   \label{longitude}
\includegraphics[width=0.6\columnwidth]{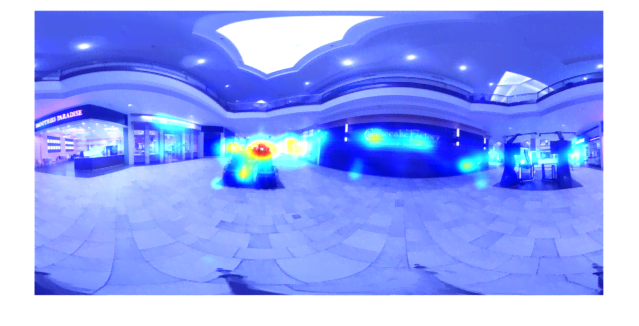}}
\subfigure[Gaussian Noise]{
   \label{latitude}
\includegraphics[width=0.6\columnwidth]{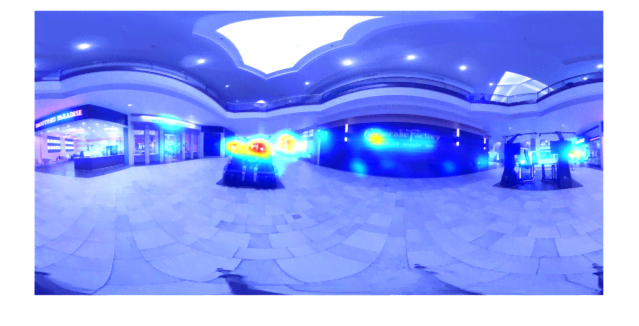}} 
\subfigure[Downsampling]{
   \label{longitude}
\includegraphics[width=0.6\columnwidth]{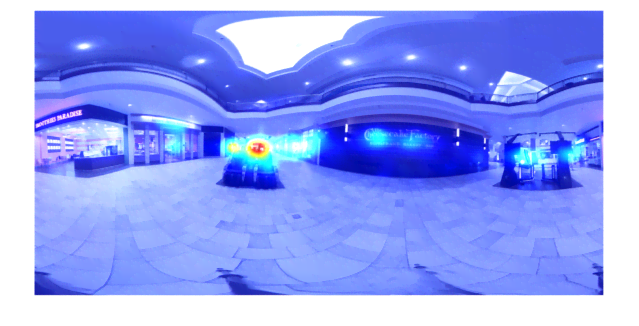}}
}
\centerline{
\subfigure[Stitching Distortion]{
   \label{longitude}
\includegraphics[width=0.6\columnwidth]{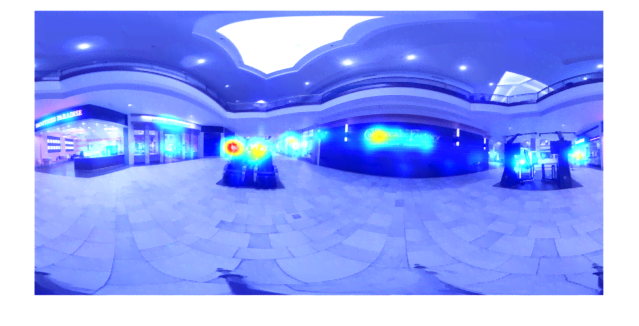}}
\subfigure[VP9]{
   \label{longitude}
\includegraphics[width=0.6\columnwidth]{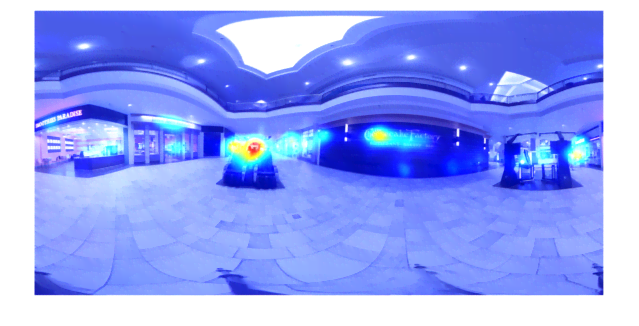}}
\subfigure[H.265]{
   \label{longitude}
\includegraphics[width=0.6\columnwidth]{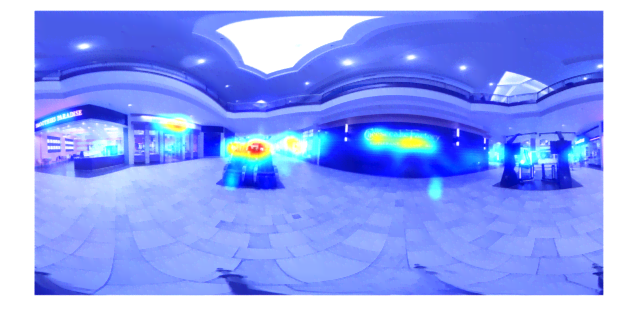}}
}

\caption{Example gaze maps on different distorted versions of a same content.}
\label{fig:gaze_content}
\end{figure*}

We calculated gaze maps using the eye tracking data recorded by the Tobii Pro. To do so, we added all gaze points for the same content, treated each as an impulse, and smoothed them by applying a Gaussian function with standard deviation of 3.34\textdegree{}\cite{rai2017dataset}. The computed gaze maps are plotted in Figure \ref{fig:db_gazemaps}. We also calculated the distribution of viewing direction for all images, as shown in 
Figure \ref{fig:latitude_longitude}. To visualize 
the distributions of viewing direction with regards to the considered distortions, exemplar plots for four of the contents are shown in Figure \ref{fig:latcontent_gazemaps} and Figure \ref{fig:longcontent_gazemaps}. Example gaze maps on different distortions of the same content are also shown in Figure \ref{fig:gaze_content}.

\subsection{Discussion of Results}
From Table \ref{srocc_table}, \ref{plcc_table} and \ref{rmse_table}, we can conclude that among all methods tested, GMSD generally performed the best while NIQE performed the worst. While WS-PSNR seems to perform better with respect to PLCC on stitching distortions, from Table \ref{f-test}, we may see that GMSD provided better quality predictions overall as compared to all other models. WS-PSNR rewards locality, hence its good performance on the stitching distortions. Since stitching distortion is highly local and it greatly affects the overall score, the deviation pooling used in GMSD is more efficient in capturing it than methods using average pooling. In addition, stitching distortion adds weak edges, which can be detected using the gradient map of images. Though MDSI also uses deviation pooling, it utilizes a fused gradient similarity map which is less efficient in detecting weak edges. As a result, it did not perform as well. From the scatter plots, it is interesting to notice that for several algorithms, the correlations for stitching distortions were very poor. This might be because of the locality property of stitching distortion that makes it more difficult. It was also interesting that both WS-PSNR and S-SSIM performed better than their counterparts, which means that applying a reprojection weight to modify traditional IQA methods can help their performance on VR images. Overall, GMSD was statistically superior to all of the other compared methods, while NIQE was statistically inferior to almost all of the others. Training on the subject data of these 3D VR images was an important step of the NR models to capture the unique perceptual peculiarities of the distorted VR image viewing experience. This is reinforced by the wide disparity in performance between the trained BRISQUE model and the training-free NIQE model, since they use the identical set of features!

From Figure \ref{fig:db_gazemaps} and \ref{fig:latitude_longitude} , we can also conclude that there exists an equator bias when viewing VR images. Subjects were more likely to view the center of the image (center bias), but this also depended on the content and whether there were objects of interest near the center. A good example is Figure \ref{gaze_e}, where the subjects' gaze was more attracted to the person in the image than to the building, although it is located at the center of the image. From Figure \ref{fig:latcontent_gazemaps}, we can see that on the various considered distortions, the distributions of the latitude viewing directions all followed the equator bias. But from Figure \ref{fig:longcontent_gazemaps}, it may be observed that this was not usually the case for the longitude viewing directions. For all of the considered distortions, the distributions tended to follow a similar trend, but on specific local distortions, the directions of interests might shift, as shown in Figure \ref{longitude_cheese}. 
By comparing the gaze maps of Figure \ref{longitude_cheese} with Figure \ref{fig:gaze_content}, we can see that the areas of interests shifted when stitching distortion was present. The appearance of stitching artifacts is much more localized as compared to other distortions. 
\section{Conclusion and Future Work}
We have created a comprehensive 3D  immersive image database with 15 different contents and 6 distortion categories rated by 40 subjects. This database is the first to evaluate the gaze-tracked quality of stereoscopic 3D VR images in an immersive environment. We also evaluated the performance evaluation of eleven popular image quality assessment algorithms. As with all LIVE Databases\cite{livedb}, the new LIVE 3D VR IQA Database is being made publicly and freely available for others to develop improved 2D and 3D VR IQA algorithms. Future work will focus on the use of  visual saliency models using the eye tracking data provided with this database, as well as developing algorithms that target VR-specific distortions like stitching, ghosting, and vignetting. 
\label{section6} 
\appendices



\ifCLASSOPTIONcaptionsoff
  \newpage
\fi


\bibliographystyle{IEEEtran}
\bibliography{IEEEexample}{}




\end{document}